\documentclass[a4paper,12pt]{article}

\usepackage{amsmath,epsfig,eucal,cite}\psfull
\newcommand{\as}{\alpha_{\mathrm{s}}}
\newcommand{\bare}{{(0)}}
\newcommand{\cut}{_{\rm cut}}
\newcommand{\dJ}[1]{\frac{\dif\sigma}{\dif J}{}\raisebox{1.5ex}{$#1$}}
\newcommand{\dif}{{\rm d}}
\newcommand{\dk}{\dif\kk}
\newcommand{\dug}{\,\raisebox{0.37pt}{:}\hspace{-3.2pt}=}
\newcommand{\e}{\varepsilon}
\newcommand{\esp}[1]{{\rm e}^{#1}}
\newcommand{\fz}{f^{(0)}}
\newcommand{\half}{\mbox{\small $\frac{1}{2}$}}
\newcommand{\hz}{h^{(0)}}
\newcommand{\id}{\boldsymbol{1}}
\newcommand{\imp}{\Longrightarrow}

\newcommand{\kk}{{\boldsymbol k}}
\newcommand{\lb}{{\boldsymbol l}}
\newcommand{\M}{{\CMcal M}}
\newcommand{\MS}{\overline{\rm MS}}
\newcommand{\mbf}[1]{{\boldsymbol #1}}
\newcommand{\mscr}[1]{\mbox{\scriptsize{$#1$}}}
\newcommand{\N}{{\CMcal N}}
\newcommand{\Nf}{N_{\!f}}
\newcommand{\ord}{{\CMcal O}}
\newcommand{\PP}{{\CMcal P}}
\newcommand{\pa}{a}
\newcommand{\pb}{b}
\newcommand{\pc}{c}
\newcommand{\pg}{g}
\newcommand{\pic}{\small}
\newcommand{\pp}{\boldsymbol p}
\newcommand{\pq}{q}
\newcommand{\qq}{{\boldsymbol q}}
\newcommand{\Sj}[1]{{\CMcal S}_J^{(#1)}}
\newcommand{\slarga}[2]{\raisebox{-#1mm}{\rule{0pt}{#2mm}}}
\newcommand{\sr}{\hat{s}}
\newcommand{\ugd}{=\hspace{-3.2pt}\raisebox{0.37pt}{:}\hspace{3pt}}

\newcommand\jhep[3]{{\it JHEP }{\bf #1} (#2) #3}
\newcommand\jp[3]{{\it J.~Phys. }{\bf #1} (#2) #3}
\newcommand\npb[3]{{\it Nucl. Phys. }{\bf B #1} (#2) #3}
\newcommand\npbps[3]{{\it Nucl. Phys. }{\bf B} { \it(Proc. Suppl.)}{ \bf #1} (#2) #3}
\newcommand\plb[3]{{\it Phys. Lett. }{\bf B #1} (#2) #3}                   
\newcommand\prd[3]{{\it Phys. Rev. }{\bf D #1} (#2) #3}            
\newcommand\zpc[3]{{\it Z. Physik }{\bf C #1} (#2) #3}             
\newcommand\sjnp[3]{{\it Sov. J. Nucl. Phys. }{\bf #1} (#2) #3}    
\newcommand\jetp[3]{{\it Sov. Phys. JETP }{\bf #1} (#2) #3}        
\newcommand\jetpl[3]{{\it JETP Lett. }{\bf #1} (#2) #3}            
\newcommand{\hep}[1]{{\tt hep-ph/#1}}

\begin{document}
\noindent
DESY 01-221   \hfill ISSN 0418--9833 
\vspace{10mm}
\begin{center}
{\Large \bf The NLO Jet Vertex for Mueller-Navelet and Forward Jets: the Quark Part}
\\[1cm]
J.\ Bartels\footnote[1]{Supported by the TMR Network ``QCD and Deep Structure of Elementary
  Particles''},
D.\ Colferai\footnote[2]{Supported by the Alexander von Humboldt Stiftung},
and G.P.\ Vacca\footnotemark[1]
\\[0.7cm]
{\small\sl
  II. Institut f\"ur Theoretische Physik; Universit\"at Hamburg\\
  Luruper Chaussee 149, 22761 Hamburg, Germany \\}
\begin{abstract}
\noindent
We calculate the next-to-leading corrections to the jet vertex which is relevant for the
Mueller-Navelet-jets production in $p\bar{p}$ collisions and for the forward jet cross
section in $ep$ collisions. In this first part we present the results of the vertex for an
incoming quark. Particular emphasis is given to the separation of the collinear divergent
part and the central region of the produced gluon.
\end{abstract}
\end{center}

\section{Introduction}

The BFKL Pomeron~\cite{BFKL76} presents the perturbative QCD prediction for the Pomeron, and
in recent years attempts have been made to verify its relevance for experimental data. Apart
from the $\gamma^*\gamma^*$ total cross section in $e^+e^-$ scattering which is generally
considered to be the gold-plated BFKL measurement~\cite{ee}, special jet measurements have
been proposed both for hadron-hadron colliders (Mueller-Navelet jets~\cite{MuNa87}) and for
deep inelastic scattering (forward jets~\cite{Mu90}). First comparisons of the leading order
calculations~\cite{2jetLL,FjetLL} with experimental data have clearly demonstrated the need
of next-to-leading order calculations: both in the $e^+e^-$ measurements at LEP and in DIS
at HERA the data are below the leading $\log s$ (LL) curves, while data from the $p\bar p$
collider at TEVATRON are found above the LL estimates.  The next-to-leading (NLL)
corrections to the BFKL kernel~\cite{FaLi98,CaCi98} lower the theoretical prediction, but
they are so large that they might even cause serious problems for the stability of the
series. Various attempts~\cite{CiCoSa99,BFKLP99} have been made in order to improve the
predictivity of the NLL BFKL approach.  However, so far one has not been able to perform a
consistent NLL analysis of the data. First of all, a consistent NLL framework for describing
not fully inclusive processes, such as jet observables, has not been established yet. In
addition, even adopting the LL high energy factorization formulae at NLL level, a few
important pieces of the NLL calculations are still missing.

\begin{figure}[ht!]
\centering
\includegraphics[width=150mm]{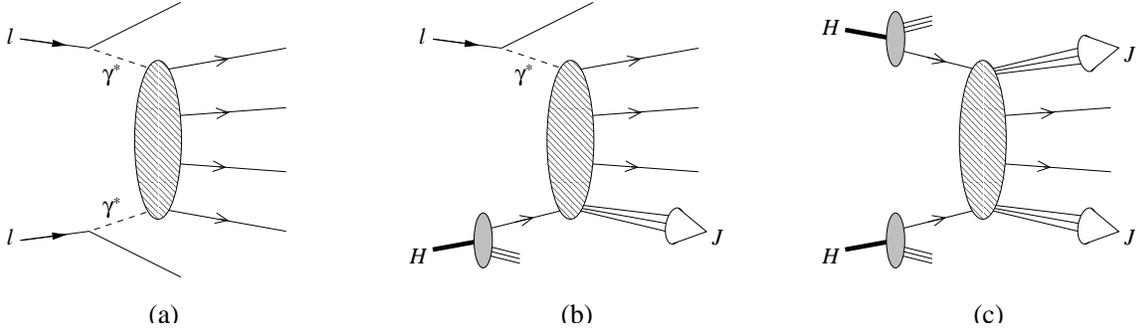}
\caption{\label{f:process}Schematic diagrams representing the three processes searching for
  high energy QCD dynamics: (a) double DIS or $\gamma^* \gamma^*$ scattering; (b) DIS with
  forward jet; (c) hadron-hadron scattering with dijets or Mueller-Navelet jets.}
\end{figure}

The three measurements for searching high energy QCD dynamics are illustrated in
Fig.~\ref{f:process}: for a complete NLL analysis one needs, in addition to the NLL
calculation of the BFKL kernel, the photon impact factor and the jet production vertex.
Whereas the former one is currently being investigated by two groups~\cite{BaGiQi00,FaMa99},
the latter one, so far, has not been calculated.  It is the purpose of this paper to present
a consistent factorization formula for high energy jet cross sections at NLL level, and to
obtain first results for the jet vertex, namely the quark-initiated vertex.  The case of an
incoming gluon will be presented in a forthcoming paper~\cite{BaCoVa02}.

The theoretical challenge in performing a NLL calculation of such a jet vertex is that it
lies at the interface of collinear factorization and BFKL dynamics: in Fig.~\ref{f:process}c
the lower incoming parton emerges from the proton and produces a hard jet, thus obeying the rules
of DGLAP evolution~\cite{DGLAP77} and collinear factorization.  Above the vertex, on the
other hand, one requires a large separation in rapidity between the two outgoing jets (which
are assumed to have large transverse energies of comparable magnitude). The kinematics
between the jets, therefore, belongs to large-$\log s$ dynamics and is described in terms of
the BFKL language.  When computing next-to-leading (NLO) corrections%
\footnote{Note the difference between LL (leading $\log s$) and LO (lowest order) and that
  between NLL (next-to-leading $\log s$) and NLO (next-to-leading order = next-to-lowest
  order = one-loop).}
to the jet production vertex, one expects to find collinear divergencies which have to be
counted as higher order corrections of the incoming parton density; at the same time, part
of the NLO corrections will overlap with high energy gluon radiation between the two jets
which belongs to the leading $\log s$ (LL) BFKL approximation.  It is one of the main goals
of our analysis to show that both types of contributions can successfully be identified and
separated from the NLL jet vertex.

As the central part of our calculation we will compute, to order $\alpha_s^3$, the high
energy limit of the cross section of the processes $\pq+\pq\to\pq+X+{\it jet}$
(Fig.~\ref{f:jet}), where $X$ may contain one gluon or quark.  The LL approximation of the
order $\ord(\alpha_s^3 \log s/s_0)$ has been calculated before~\cite{FoRo97}, we will
present the NLL (constant in $s$) term $\ord(\alpha_s^3)$.  We will show that the cross
section can be written in a factorized form: there are NLL corrections to the impact factor
of the upper incoming quark (which have been calculated before~\cite{Ci98}), and for the
emission of the gluon in the central region we recover the LL BFKL result. Finally, the NLL
corrections to the jet vertex of the lower incoming quark are what we obtain as new result.
Making use of this high energy factorizing, we can use our results for the
Mueller-Navelet jets (Fig~\ref{f:process}c): we can apply the NLL results not only to the
lower jet vertex but also to the upper one. For the forward jets in DIS
(Fig.~\ref{f:process}b) we have the NLL corrections for the jet vertex, but we have to wait
for the completion of the NLL corrections to the photon impact factor.

An important result of the NLL calculation is the dependence upon the energy scale $s_0$: in
the leading $\log s$ approximation this scale is undetermined and thus introduces a principal
uncertainty of the theoretical prediction.  The NLL calculation determines how the cross
section changes with a change in $s_0$ and removes this uncertainty.

Moreover, at next-to-leading order, the renormalized parton densities start to play a role,
reducing drastically the dependence on the collinear factorization scale $\mu_F$, otherwise
maximal in all previous LO calculations.

Our paper will be organized as follows. We begin with a brief outline of our program. In
Secs.~\ref{s:virt} and \ref{s:rc} we present the results of our calculation: first the
virtual corrections, then the real corrections. Our main emphasis will be on the separation
of soft and collinear divergencies at the vertex of the lower incoming quark and on the
removal of the central region of the produced gluon. Combining the real and virtual
corrections we obtain in Sec.~\ref{s:jv} an analytic expression for the jet vertex.
We conclude with a brief summary and discussion.

\section{High energy factorization\label{s:hef}}

\subsection{General framework\label{s:gf}}

\begin{figure}[hb!]
\centering
\includegraphics[width=60mm]{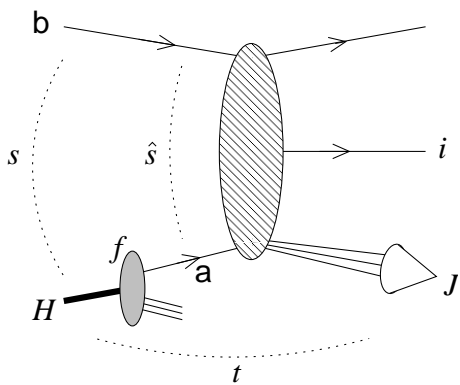}
\caption{\label{f:jet}Diagrammatic representation of the high energy process studied in this
  paper. $H$ is the incoming hadron providing parton $\pa$ with distribution density $f$;
  $\pb$ is the other incoming particle which will be taken to be a quark; $J$ denotes the
  jet produced in the forward direction (w.r.t $H$) and $i$ is the generic label for
  outgoing particles.}
\end{figure}

The processes that we are going to study are those in which a hadron $H$ strongly interacts
with parton $\pb$; to be definite, we chose $\pb$ to be a quark. In the final state a jet
$J$ (in the forward direction with respect to $H$) is then identified (see
Fig.~\ref{f:jet}). Our notation uses light cone coordinates
\begin{equation}\label{lcc}
 p^\mu=(p^+,p^-,\pp)\;,\quad p^\pm\dug\frac{p^0 \pm p^3}{\sqrt2}\;,
\end{equation}
where the light-like vectors $p_H$ and $p_\pb$ form the basis of the longitudinal plane:
\begin{subequations}
\begin{align}
 \label{coord1}
 p_H &= \left(\sqrt{\frac{s}{2}},0,\mbf{0}\right)\;,\quad s\dug(p_H+p_\pb)^2\\
 \label{coord2}
 p_\pb &= \left(0,\sqrt{\frac{s}{2}},\mbf{0}\right)\\
 \label{coord3}
 p_i &= E_i\left(\frac{\esp{y_i}}{\sqrt2},\frac{\esp{-y_i}}{\sqrt2},\mbf{\phi}_i\right)\;.
\end{align}
\end{subequations}
In the last equation we have introduced a parameterization for the $i$-th particle in the
final state in terms of the rapidity $y_i$ (in the $p_H+p_\pb$ center of mass frame), of
the transverse energy $E_i=|\pp_i|$ and of the azimuthal unit vector
$\mbf{\phi}_i\parallel\pp_i$.

As usual, our jet consists of a certain number of partons, 
whose rapidities and azimuthal angles are
found inside a given (small) region in the $(y,\phi)$ plane. The position of the center of
that region defines the rapidity $y_J$ and the azimuthal angle $\phi_J$ of the jet, its size
is related to the jet radius $R$, and the sum of the transverse energies of the
particles forming the jet constitutes the jet transverse energy $E_J$. We will 
keep our jet definition rather general; it only has to obey a few minimal 
requirements (see Sec.~\ref{s:jd}).

We will be interested in the high energy limit of the subprocess 
$\pa+\pb\to\pq+i+{\it jet}$. In addition to the energy $s$, we need to define 
the momentum transfer
\begin{equation}\label{deft}
 t \dug (p_H-p_J)^2\;.
\end{equation}
and the jet energy $E_J$:
\begin{equation}\label{valt}
 p_J \simeq E_J\left(\frac{\esp{y_J}}{\sqrt2},\frac{\esp{-y_J}}{\sqrt2},\mbf{\phi}_J\right)
 \quad\imp\quad -t\simeq\sqrt{s}E_J\esp{-y_J}\;.
\end{equation}
The high energy (Regge) limit that we are considering is defined by
\begin{equation}\label{HElimit}
 E_J^2\;\text{fixed}\;,\quad t\;\text{fixed}\;,\quad s\to\infty
\end{equation}
which determines a logarithmic growth of the jet rapidity with $s$:
\begin{equation}\label{logrowth}
 y_J\sim\half\log s\;.
\end{equation}
The use of perturbation theory is justified because we consider all kinematic scales much
greater than the QCD scale:
\begin{equation}\label{hard}
 E_J^2\sim-t\gg\Lambda_{\rm QCD}^2;.
\end{equation}
This condition allows also us to neglect the masses and the Fermi motion of the light
partons inside $H$. Equivalently, in terms of the twist expansion, the leading twist
contribution can be extracted by considering the parton massless and collinear to $H$. We
shall therefore adopt
\begin{equation}\label{defpa}
 p_\pa = x\,p_H =\left(x\sqrt{\frac s2},0,\mbf{0}\right)\;.
\end{equation}
 
It is well known that the Regge limit is dominated by gluon $t$-channel
exchanges~\cite{BFKL76} and that, in the leading logarithmic approximation, the elastic
scattering amplitude and the total cross section can be written in a factorizing form:
{\it (i)} a gluon Green's function which describes the exchanged system, and {\it (ii)} impact
factors which denote the coupling to the scattering partons or particles. NLO corrections
that have been computed for the BFKL kernel~\cite{FaLi98,CaCi98} and for a few impact
factors~\cite{Ci98,CiCo98,FFKP00,CiRo00} support this factorizing form also in
next-to-leading order.  It is one of the goals of this paper to verify this factorization
also for the inelastic process $\pa+\pb\to\pa+i+{\it jet}$. Thanks to this property we will be able
to treat, on the same footing, several classes of processes.  Two of them are crucial for
the study of QCD in the Regge limit:
\begin{itemize}
\item {\it dijet}, or Mueller-Navelet jets, coming from hadron-hadron collisions where a jet
  is detected in the forward direction of each hadron; in this case particle $\pb$ is a
  hadron;
\item {\it forward jet}, coming from lepton-hadron collisions where a jet is detected in the
  forward direction with respect to the hadron; in this case particle $\pb$ is identified
  with the virtual gauge boson, e.g. a photon, emitted by the lepton.
\end{itemize}

In the following we assume that a proper definition of the jet has been chosen. This choice
is represented by a function (actually a distribution) $S_J$ which selects the final state
configurations contributing to the observable we are interested in. 
The jet cross section is given by the
action of $S_J$ on the full exclusive cross section in $D=4+2\e$ spacetime dimensions
\begin{equation}\label{exclusive}
 \dif\sigma = \frac{1}{2s}\sum_{n=2}^\infty (2\pi)^D
 \delta^D(p_H+p_\pb-\sum_{i=2}^n p_i)\langle|\M_{H\pb\to n}|^2\rangle
 \,\dif\Phi_n(p_1,\cdots,p_n)
\end{equation}
as follows:
\begin{equation}\label{dsigmaJ}
 \dJ{} \dug \frac{\dif\sigma}{\dif y_J \dif E_J \dif \phi_J} = \int\;\dif\sigma\,S_J\;.
\end{equation}
Here $J=(y_J,E_J,\phi_J)$ collects the jet variables, $n$ is the number of particles in the
final state, $\M$ is the invariant amplitude and $\dif\Phi$ is the phase space measure.

In order to describe perturbatively a hadron-initiated process, we assume --- according to
the parton model --- the physical cross section to be given by the corresponding partonic
cross section $\dif\hat\sigma_\pa$ (computable in perturbation theory) convoluted with the
distribution densities $f_\pa$ of the partons $\pa$ inside the hadron $H$. The partonic
distribution functions (PDFs) $f_\pa:\pa\in H$ constitute a non-perturbative input. This
approach is justified provided the infrared singularities stemming from QCD interaction
among massless objects can be consistently absorbed in a redefinition of the PDFs according
to the well known factorization of mass singularities~\cite{CoSoSt89}. Those ``renormalized
PDFs'' will be eventually interpreted as the universal objects measured in hadronic
collisions and obeying the DGLAP equations. Let us therefore write
\begin{equation}\label{collfact}
 \dif\sigma = \sum_{\pa\in H}\int_0^1\dif x\;\dif\hat\sigma_{\pb\pa}(x)\,\fz_\pa(x)
\end{equation}
where $x=p_\pa^+/p_H^+$ is the longitudinal momentum fraction of the parton $\pa$ with
respect to the parent hadron $H$. In Eq.~(\ref{collfact}) we show explicitly that, according
to the previous discussion, the PDFs must still be considered as ``bare'' quantities.  In
conclusion, the jet cross section is given by
\begin{equation}\label{pSf}
 \dJ{} = \sum_\pa\int\dif x\;\dif\hat\sigma_{\pb\pa}(x)\,S_J(x)\fz_\pa(x)
\end{equation}
and is diagrammatically represented in Fig.~\ref{f:jet}.

We proceed by reviewing parton-hadron scattering at lowest order.

\subsection{The Jet vertex at lowest order\label{s:LO}}

\begin{figure}[b]
\centering
\includegraphics[height=30mm]{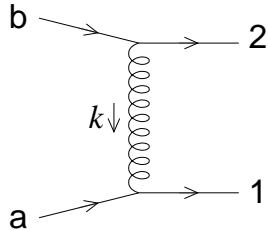}
\caption{\label{f:ab12}Leading diagram at lowest order for parton-parton scattering: the
  interaction occurs via gluon exchange in the $t$-channel.}
\end{figure}

In order to evaluate the jet cross section in the high energy regime~(\ref{HElimit}) for
parton-hadron scattering, thanks to Eq.~(\ref{pSf}), we need only to consider parton-parton
scattering. At lowest order (LO), the relevant cross section is dominated by one gluon
exchange in the $t$-channel, as shown in Fig.~\ref{f:ab12}.
Let us define the gluon momentum and the partonic center of mass energy squared
\begin{align}\label{defk}
 k &\dug p_1-p_\pa = -z p_\pa+w p_\pb + k_\perp\;,\quad k_\perp = (0,0,\kk)\;,
 \\ \nonumber \sr &\dug (p_\pa+p_\pb)^2 = xs \;.
\end{align}
The partonic cross section is constant in $\sr$ and is given by
\begin{equation}\label{LOpart}
 \dif\hat\sigma^{(0)}_{\pb\pa} =
 \hz_\pb(\kk)\hz_\pa(\kk)\dif\kk\;,\quad\dif\kk\equiv\dif^{2+2\e}\kk\;,
\end{equation}
in terms of the LO partonic impact factors~\cite{CiCo98}
\begin{equation}\label{dh0}
 \hz_i(\kk) \dug \N\,\frac{C_i}{\kk^2}\;,\quad 
 \N = \frac{2^{1+\e} \as}{\mu^{2\e}\Gamma(1-\e)\sqrt{N_c^2-1}}\;,\quad (i=\pa,\pb)\;,
\end{equation}
where the colour factor $C_i$ is $C_F=(N_c^2-1)/2N_c$ for a quark ($i=\pq$) and $C_A=N_c$
for a gluon ($i=\pg$); the coupling $\as$ in $D=4+2\e$ dimensions is defined in
Eq.~(\ref{defas}).

It is evident that the jet can contain only one of the two particles in the final state, the
other moving in the opposite direction. Furthermore, since we are looking 
for the jet in the
forward direction with respect to $H$, the configuration $p_2=p_J$ gives a negligible
contribution to the cross section. This can be easily seen by noticing that the
corresponding amplitude involves a propagator $\sim 1/|u| \simeq 1/s$ much smaller than that
of the $p_1=p_J$ amplitude $\sim 1/t$. This means that, at lowest order, the jet momentum
has to be identified with $p_1$, so that the jet distribution for two-particle final states,
according to Eq.~(\ref{dsigmaJ}), reads
\begin{equation}\label{j2}
 S_J^{(2)}(p_1,p_2;p_\pa,p_\pb) = \delta(y_1-y_J)\delta(E_1-E_J)\delta(\phi_1-\phi_J)\;.
\end{equation}
As independent variables in $S_J^{(2)}$ we can adopt $\pp_1=\kk$ for the final state, and
$p_\pa^+ / p_H^+ = x$ for the initial state, so that we can define
\begin{equation}\label{dS2}
  \Sj2(\kk;x) \dug S_J^{(2)}(p_1,p_2;p_\pa,p_\pb)
 = \delta\left(1-\frac{x_J}{x}\right)E_J^{1+2\e}\delta(\kk-\kk_J)\;,\quad
 x_J \dug \frac{E_J\esp{y_J}}{\sqrt s}\;,
\end{equation}
being $\kk_J$ the transverse momentum of the jet. By substituting Eqs.~(\ref{LOpart}) and
(\ref{dS2}) in Eq.~(\ref{pSf}), the LO jet cross section assumes the factorized form
\begin{equation}\label{fattLO}
 \dJ{^{(0)}} = \sum_\pa \int\dif x \int\dif\kk\;\hz_\pb(\kk)\hz_\pa(\kk)\Sj2(\kk;x)\fz_\pa(x)\;.
\end{equation}
Besides the PDF $f_\pa$ and the partonic impact factor $h_\pb$, we are left with a term
that can be interpreted as the LO jet vertex
\begin{equation}\label{dV0}
 V^{(0)}_\pa(\kk,x) \dug \hz_\pa(\kk)\Sj2(\kk;x)\;.
\end{equation}
The lowest order formula for the jet cross section is therefore
\begin{equation}\label{LOFF}
 \dJ{^{(0)}} = \sum_\pa \int\dif x \int \dif\kk\;\hz_\pb(\kk)V^{(0)}_\pa(\kk,x)\fz_\pa(x)\;.
\end{equation}

\subsection{One-loop analysis: LL approximation and future strategy\label{s:1la}}

Moving on to higher order, let us first address the leading logarithmic approximation, i.e.,
terms of the order $\as^n\log^n(s/-t)$.  The resummation of these terms, also referred to as
leading lo\-ga\-rithmic (LL) approximation, was addressed long ago for fully inclusive
processes~\cite{BFKL76}, and it has also been applied to dijets~\cite{2jetLL} and forward
jets~\cite{FjetLL}. It is instructive to review briefly how the logarithmic enhanced terms
arise at one-loop, and to study the structure of the singularities and their connection with
the various kinematic regions.

\begin{figure}[ht!]
\centering
\includegraphics[height=30mm]{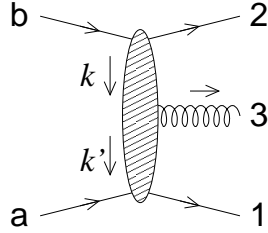}
\caption{\label{f:ab123}Labelling the two-to-three parton scattering process. Note the
  definition of the transferred momenta $k$ and $k'$.}
\end{figure}

We consider first the real corrections to, say, quark-quark scattering, which involves the
emission of an additional gluon of momentum $p_3$ as shown in Fig.~\ref{f:ab123}. The
structure of the final states giving the leading contributions corresponds to the so called
multi-Regge kinematics (MRK)
\begin{equation}\label{MRK}
 y_1 \gg y_3 \gg y_2\;,\quad
 E_1 \sim E_3 \sim E_2\;,
\end{equation}
where the rapidity $y_3$ of the emitted gluon is strongly ordered between the rapidities of
the scattered partons $y_1$ and $y_2$ (i.e., the gluon is emitted in the central region),
while the magnitudes of the transverse momenta are of the same order.
\begin{figure}[hb!]
\centering
\includegraphics[width=140mm]{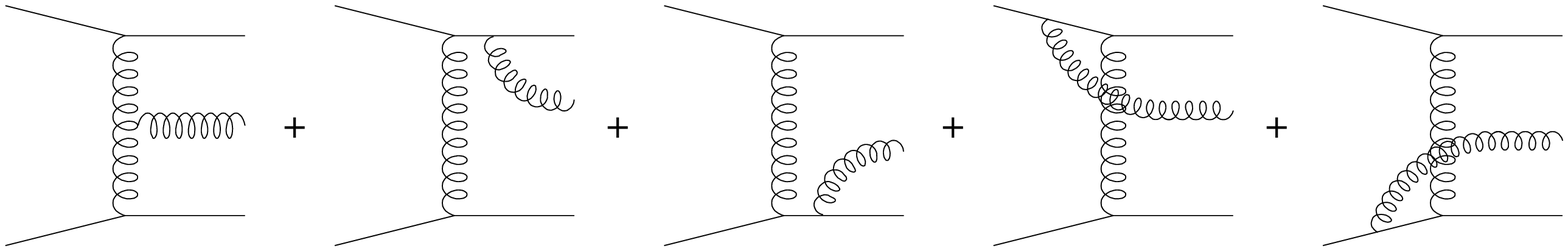}
\caption{\label{f:magnifici5}Feynman diagrams contributing to the $\pq\pq\to\pq\pq\pg$
  process at leading $\log s$ order.}
\end{figure}

All diagrams shown in Fig.~\ref{f:magnifici5} contribute at LL, and the resulting
differential partonic cross section reads
\begin{equation}\label{central}
 \dif\hat\sigma_{\pb\pa}^{(1,\text{real})} \simeq \as\hz_\pb(\kk)\hz_\pa(\kk')
 \frac{C_A}{\pi}\frac1{\pi_\e(\kk-\kk')^2}\,\dif\kk\,\dif\kk'\,\frac{\dif z}{z}\;,
\end{equation}
where $\pi_\e$ is given in Eq.~(\ref{Pgq}). We have introduced the momenta transferred by
the $\pa$ and $\pb$ partons
\begin{subequations}\label{k}
\begin{align}\label{k1}
 k &\dug p_\pb - p_2 = -\bar{w} p_\pa + w p_\pb + k_\perp\;,&
 k_\perp &= (0,0,\kk)\\ \label{k2}
 k' &\dug p_1 - p_\pa = -z p_\pa + \bar{z} p_\pb + k'_\perp\;,&
 k'_\perp &= (0,0,\kk')\;.
\end{align}
\end{subequations}
The transverse energies introduced in Eq.~(\ref{coord3}) correspond to
$E_1=|\kk'|,E_2=|\kk|,E_3=|\kk-\kk'|$.  In the MRK~(\ref{MRK}), $\bar w\sim\bar z\ll w\sim
z\ll 1$, so that the outgoing parton 1 carries most of the plus-component of the
parent parton $\pa$.  By using Eq.~(\ref{coord3}) and comparing with Eq.~(\ref{defpa}) we
get
\begin{equation}\label{kinLL}
 y_1 \simeq \log\frac{x\sqrt s}{E_1}\;.
\end{equation}
Since $E_1=|\kk'|$, $\kk'$ characterizes completely the momentum $p_1$.  The range of the
momentum fraction $z$ is approximately given by $\pp_3^2/\sr<z<1-\kk'{}^2/\sr$. The boundary
values of $z$ correspond to the gluon in the fragmentation region of parton $\pb$ and $\pa$
respectively, where Eq.~(\ref{central}) no longer holds. Nevertheless, it shows that, upon
$z$-integration, a logarithmic factor $\log(\sr/s_0)$ appears:
\begin{align}\label{realNLO}
 &\dif\hat\sigma^{(1,{\rm real})}_{\pb\pa} = \int\dif z\;
 \frac{\dif\hat\sigma}{\dif z}{}\raisebox{1.5ex}{$^{(1,{\rm real})}_{\pb\pa}$} 
 = \as\hz_\pb(\kk)\hz_\pa(\kk')
 \left[K^{(0,{\rm real})}(\kk,\kk')\log\frac{\sr}{s_0}
 +\text{const}\right]\,\dif\kk\,\dif\kk' \\ \label{dK0real}
 &K^{(0,{\rm real})}(\kk,\kk') \dug \frac{C_A}{\pi}\frac1{\pi_\e(\kk-\kk')^2}\;.
\end{align} 
The coefficient of the $\log s$ term $K^{(0,{\rm real})}$ is the real part of the leading
log BFKL kernel.  The {\em scale of the energy} $s_0$ is a parameter of the order of the
transverse momenta squared of the gluons
($s_0\sim\pp_3^2=|\kk-\kk'|^2\sim\kk^2\sim\kk'{}^2$). Its value is not determined in the LL
approximation --- a change of $s_0$ would affect only the constant piece in
Eq.~(\ref{realNLO}) ---, but will play a central role when going to the next-to-leading
level of accuracy. In particular, that part of the differential cross section in which the
emitted gluon lies outside the central region contributes to the constant term (i.e.,
without a $\log s$ enhancement).

\begin{figure}[hb!]
\centering
\includegraphics[width=130mm]{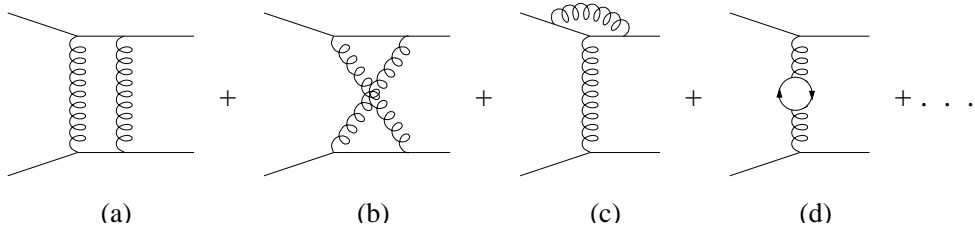}
\caption{\label{f:NLOvirt}Feynman diagrams contributing to the one-loop virtual corrections to
  $\pq\pq\to\pq\pq$ elastic scattering in LL approximation (a,b) and in NLL approximation
  (c,d,...).}
\end{figure}

Let us now consider the virtual corrections. In covariant gauges, the diagrams involving two
gluon exchanges (see Figs.~\ref{f:NLOvirt}a,b) give contributions to the amplitude which
increase logarithmically with the energy, whereas other diagrams (shown in
Figs.~\ref{f:NLOvirt}c,d) have no logarithmic enhancement.  Note, however, that the latter
diagrams contain the ultra-violet (UV) singularities that provide the renormalization of the
coupling.

The result of the virtual correction to the partonic cross section can be presented as follows:
\begin{equation}\label{virtNLO}
 \dif\hat\sigma_{\pb\pa}^{(1,\text{virt})} = \as\hz_\pb(\kk)\hz_\pa(\kk)
 \left[2\omega^{(1)}(\kk)\log\frac{\sr}{\kk^2}+\text{const}\right]\dif\kk\;.
\end{equation}
The coefficient $\omega^{(1)}$ in front of the $\log s$ is the {\em one-loop Regge-gluon
  trajectory}, because --- according to Regge theory --- the amplitude for $\pa\pb\to\pa\pb$
elastic scattering mediated by color octet exchange is described by the exchange of a
reggeized gluon and takes the form:
\begin{equation}\label{regge}
 \M_{\pb\pa} \sim \left(\frac s{-t}\right)^{\omega(t)} =
 1+\as\omega^{(1)}(t)\log\frac s{-t} +\cdots\;.
\end{equation}
By taking the square modulus and adjusting the normalization through the LO expression
(\ref{LOpart}), it is straightforward to match Eq.~(\ref{virtNLO}).  This contribution to
the cross section defines the virtual part of the LL BFKL kernel:
\begin{equation}\label{dK0virt}
 K^{(0,{\rm virt})}(\kk,\kk') \dug 2\omega^{(1)}(\kk)\delta(\kk-\kk')\;.
\end{equation}
Together with Eqs.~(\ref{central}) and (\ref{virtNLO}) we obtain the full one-loop partonic
cross section
\begin{equation}\label{NLOpart}
 \dif\hat\sigma_{\pb\pa}^{(1)} = \as\hz_\pb(\kk)\hz_\pa(\kk')
 \left[K^{(0)}(\kk,\kk')\log\frac{\sr}{s_0}+\text{const}\right]\dif\kk\,\dif\kk'\;.
\end{equation}
Here $K^{(0)}$ collects the real and virtual coefficients of the logarithmic terms:
\begin{equation}\label{K0}
 K^{(0)}(\kk,\kk') \dug K^{(0,{\rm real})}(\kk,\kk') + K^{(0,{\rm virt})}(\kk,\kk')\;.
\end{equation}
and defines the LL BFKL kernel; the term $const$ contains contributions that have no large
energy logarithm.

Let us use these results to define the jet cross section. The jet we want to observe
includes the outgoing particles carrying the largest values of rapidity lying within a
specified (small) range $R$. However, the strong rapidity order in Eq.~(\ref{MRK}) allows
only $p_1$ to enter the jet, so that the jet distribution for three particles in the final
state $S_J^{(3)}$ reduces to the lowest order one $S_J^{(2)}$ (Eq.~(\ref{j2})) which,
combined with the impact factor $\hz_\pa$ in Eq.~(\ref{NLOpart}), reproduces the lowest
order jet vertex $V_\pa^{(0)}$ (Eq.~(\ref{dV0})).  Convoluting with the PDF we obtain the
following factorized expression for the one-loop LL jet cross section ($\sr=xs$):
\begin{equation}\label{NLOFF}
 \dJ{^{(1,\rm LL)}} = \as\sum_\pa\int\dif x \int \dif\kk\,\dif\kk'\;
 \hz_\pb(\kk) K^{(0)}(\kk,\kk')\log\frac{xs}{s_0} V_\pa^{(0)}(\kk',x) \fz_\pa(x)\;.
\end{equation}

It is the purpose of this paper to compute, in this order of $\alpha_s$, the constant (i.e.,
non-logarithmic) corrections to our jet cross section formula. We generalize the
results~(\ref{LOFF}) and~(\ref{NLOFF}) and make the following {\em ansatz}:
\begin{subequations}\label{ansatz}
\begin{align}\label{FF}
 \dJ{} &= \as\sum_\pa\int\dif x \int \dif\kk\,\dif\kk'\;
 h_\pb(\kk) G(xs,\kk,\kk') V_\pa(\kk',x) f_\pa(x)\;, \\ \label{G1loop}
 G(xs,\kk,\kk') &\dug \delta(\kk-\kk')+\as K^{(0)}(\kk,\kk')\log\frac{xs}{s_0}\;,
\end{align}
\end{subequations}
According to the previous remarks, we suppose that the inclusion of the one-loop constant
terms just provides perturbative corrections to the quark impact factor, to the jet vertex,
and to the PDF as follows:
\begin{subequations}\label{svil}
\begin{align}\label{svilh}
 h &= \hz + \as h^{(1)} \\ \label{svilV}
 V &= V^{(0)} + \as V^{(1)} \\ \label{svilf}
 f &= \fz + \as f^{(1)}\;.
\end{align}
\end{subequations}
Equivalently, our ansatz corresponds to the following structure for the one-loop cross
section:
\begin{align}\label{dJ1loop}
 \dJ{^{(1)}} &= \as\sum_\pa\int\dif x \int \dif\kk\; \bigg\{ 
 \int\dif\kk'\left[\hz_\pb(\kk) K^{(0)}(\kk,\kk')\log\frac{xs}{s_0} 
 V_\pa^{(0)}(\kk',x) \fz_\pa(x) \right] + \\ \nonumber 
 &\quad\; h^{(1)}_\pb(\kk)  V_\pa^{(0)}(\kk,x) \fz_\pa(x)
 + \hz_\pb(\kk)  V_\pa^{(0)}(\kk,x) f^{(1)}_\pa(x)
 + \hz_\pb(\kk)  V_\pa^{(1)}(\kk,x) \fz_\pa(x) \bigg\}\;,
\end{align}
\noindent
which is obtained simply by expanding Eq.~(\ref{FF}) up to relative order $\as$.

In Eq.~(\ref{dJ1loop}) the Born approximations (marked by the superscript $(0)$) have been
listed in~(\ref{dh0}), (\ref{K0}), and (\ref{dV0}) respectively.  For the first order
correction to the partonic impact factor, $h^{(1)}$, which appears on the second line, we
can use the known expression of Ref.~\cite{Ci98,CiCo98}, and for the correction to the PDF,
$f^{(1)}$, we have the usual LO Altarelli-Parisi splitting functions (in the $\MS$ scheme):
\begin{align}\label{deff1}
 f^{(1)}_\pa(x,\mu_F^2) &\dug
 \frac1{2\pi\e}\left(\frac{\mu_F^2}{\mu^2}\right)^\e \sum_\pc\int_x^1\frac{\dif\xi}{\xi}\;
 P_{\pa\pc}(\xi) \fz_\pc\left(\frac{x}{\xi}\right) \\ \nonumber
 &= \frac1{2\pi\e}\left(\frac{\mu_F^2}{\mu^2}\right)^\e 
 \sum_\pc P_{\pa\pc}\otimes \fz_\pc\;.
\end{align}
Finally, the correction term $V_\pa^{(1)}$ is what we want to compute in this paper.

Eqs.~(\ref{ansatz}) and (\ref{svil}) constitute a highly non trivial ansatz, which will be
shown to depend upon a careful separation of singular and finite pieces. Our main task will
consist to identify the collinear singularities~(\ref{deff1}) which go into the
renormalization of the parton densities, to check that the other infrared singularities
cancel out when adding virtual and real corrections, and, finally, to separate the terms
proportional to $\log s$ which belong into the first line of~(\ref{dJ1loop}). At the end, a
finite term remains, which eventually can be interpreted as one-loop correction to the jet
vertex, $V_\pa^{(1)}$.

In the rest of this paper we will compute this jet vertex correction, $V_\pa^{(1)}$. In this
paper we will concentrate on the case of incoming quark ($\pa=\pq$). The case of an incoming
gluon ($\pa=\pg$) will be dealt with in a subsequent paper~\cite{BaCoVa02}.

\section{Virtual corrections\label{s:virt}}

In the following we will develop the one-loop analysis of the quark-initiated jet production
process. We adopt dimensional regularization in $D=4+2\epsilon$ dimensions and define,
according to the $\MS$ scheme, the dimensionless coupling $\as$ as a function of the
dimensionful bare coupling $g$ and of the renormalization scale $\mu$ as follows:
\begin{equation}\label{defas}
 \as = \as^\bare \dug \frac{g^2 \mu^{2\e}\Gamma(1-\e)}{(4\pi)^{1+\e}}   
\end{equation}

We begin by collecting the virtual corrections. Some of the diagrams are shown in
Fig.~\ref{f:NLOvirt}. Discarding all terms suppressed by powers of $s$, the one-loop
parton-parton cross section can be derived from Ref.~\cite{PPRcorr} and reads
\begin{equation}\label{Vpart}
 \dif \hat\sigma_{\pb\pa} = \as\,\hz_\pb(\kk)\hz_\pa(\kk)\left[
 2\omega^{(1)}(\kk)\log\frac{xs}{\kk^2} + \Pi_{\pb}(\kk) + \Pi_{\pa}(\kk) \right]\dk\;.
\end{equation}
The first term has already been introduced in Sec.~\ref{s:1la}: it represents the LL
contribution to the virtual corrections. In particular the coefficient of $\log s$, namely
$2\omega^{(1)}$, constitutes the virtual part of the leading kernel $K^{(0)}$ of
Eq.~(\ref{K0}) and is just twice the one-loop Regge-gluon trajectory
\begin{equation}\label{traj}
\omega^{(1)}(\kk) = -\frac{C_A}{\pi}\frac{1}{2\e}\frac{\Gamma^2(1+\e)}{\Gamma(1+2\e)}
\left( \frac{\kk^2}{\mu^2}\right)^\e\;.
\end{equation}
It shows an $\e$-pole due to a soft singularity which compensates the corresponding one of
the real part of the kernel.

The non logarithmic terms in Eq.~(\ref{Vpart}) represent the NLL contribution to the virtual
corrections and are expressed in terms of the virtual corrections to the impact factor
$\Pi$.  The virtual corrections to the quark impact factor read:
\begin{equation}\label{qIF}
 \Pi_{\pq}(\kk) = \left[
 -\frac{11N_c - 2\Nf}{12\pi}\frac{1}{\e}
 +\left(\frac{85}{36}+\frac{\pi^2}{4}\right)\frac{C_A}{\pi}-\frac{5}{18}\frac{\Nf}{\pi}
 -\left(\frac{1}{\e^2}-\frac{3}{2\e}+4-\frac{\pi^2}{6}\right)\frac{C_F}{\pi}
 \right]\left(\frac{\kk^2}{\mu^2}\right)^\e\;.
\end{equation}
In the above expression we have singled out terms of different physical origin.  The first
term is proportional to the $\beta$-function coefficient $b_0 = (11N_c - 2\Nf)/12\pi$. It
multiplies the ultraviolet (UV) pole providing the $\MS$ renormalization of the coupling
\begin{equation}\label{run}
 \as(\kk^2) \dug \as^\bare\left[1-\as^\bare\frac{b_0}{\e}
 \left(\frac{\kk^2}{\mu^2}\right)^\e\right]\;.
\end{equation}
In fact, at LO the partonic cross section~(\ref{LOpart}) is simply the product of two bare
partonic impact factors
\begin{equation}\label{hh}
 \frac{\dif\hat\sigma}{\dk}{}\raisebox{8pt}{$^{(0)}_{\pb\pq}$} = 
 \hz_\pb(\as^\bare)\,\hz_\pq(\as^\bare)\;,
\end{equation}
where we have explicitly shown only the dependence on $\as^\bare$.  Adding the UV divergent
term of Eq.~(\ref{Vpart}) stemming from $\Pi_\pq$ renormalizes the coupling inside $h_\pq$:
\begin{equation}\label{renh}
 \frac{\dif\hat\sigma}{\dk}{}\raisebox{1.5ex}{$^{(0)}_{\pb\pq}$} +
 \frac{\dif\hat\sigma}{\dk}{}\raisebox{1.5ex}{$^{(1,\text{UV})}_{\pb\pq}$}
 \left.\frac{}{}\hspace{-0.5em}\right|_{\Pi_{\pq}} = \hz_\pb(\as^\bare)\,\hz_\pq(\as^\bare)
 \left[1-\as^\bare\frac{b_0}{\e}
 \right]
 = \hz_\pb(\as^\bare)\,\hz_\pq(\as(\mu^2))\;. 
\end{equation}
The same UV pole can be found in $\Pi_{\pb}$, and it provides the running of the coupling of
$\hz_\pb$.

The virtual contribution to the jet cross section is easily obtained by substituting in
Eq.~(\ref{pSf}) the expression~(\ref{dS2}) for the jet distribution and Eq.~(\ref{Vpart})
for the partonic cross section. By combining the jet distribution with the quark $\pq$
impact factor at LO we reproduce the LO jet vertex~(\ref{dV0}) and, after renormalization,
we end up with
\begin{equation}\label{dJvirt}
 \dJ{^{(\rm virt)}} = \as\int\dif x\int\dk\;\hz_\pb(\kk)
 \left[2\omega^{(1)}(\kk)\log\frac{xs}{\kk^2}
 +\widetilde\Pi_\pb(\kk)+\widetilde\Pi_\pq(\kk)\right]
 V^{(0)}_\pq(\kk,x) \fz_\pq(x)\;,
\end{equation}
where
\begin{align}\label{renqIF}
 \widetilde\Pi_{\pq}(\kk) &\dug \Pi_{\pq}(\kk)-(-b_0/\e)\\ \nonumber
 &=\left[\left(\frac{85}{36}+\frac{\pi^2}{4}\right)\frac{C_A}{\pi}-\frac{5}{18}\frac{\Nf}{\pi}
 -\left(\frac{1}{\e^2}-\frac{3}{2\e}+4-\frac{\pi^2}{6}\right)\frac{C_F}{\pi}
 -b_0\log\frac{\kk^2}{\mu^2}\right]\left(\frac{\kk^2}{\mu^2}\right)^\e\;. 
\end{align}
Any occurrence of $\as$ in Eq.~(\ref{dJvirt}) and in all other coming formulae is to be
understood as $\as(\mu^2)$.

The ``reduced'' quark impact factor virtual correction~(\ref{renqIF}) shows double and
single poles in $\e$. These poles are of IR origin and are due to both soft and collinear
singularities. Partly they will cancel against the corresponding singularities of the real
emission corrections, leaving a simple pole that will be absorbed in the redefinition of the
PDFs. This will be shown in Sec.~\ref{s:jv}.

\section{Real corrections\label{s:rc}}

When calculating the real emission corrections to the one-loop jet cross section, our main
concern will be the correct treatment of the IR singularities (keeping in mind that we are
considering a partially exclusive process). Infrared singularities are contained in the
upper quark impact factor, the real part of the BFKL kernel, and in the lower jet vertex.
Some of the latter ones contribute to the renormalization of the incoming parton density.
{\em A priori} it is not evident that the overlap of the various regions of the phase space
responsible for the divergencies can be disentangled in such a way that they reproduce all
the expected singularities (and not more). We will show that this is actually the case.

\subsection{Jet definition\label{s:jd}}

We begin with a brief review of the jet definition. We follow the arguments given
in~\cite{jetdef}, and we wish to keep our distribution functions $S_J^{(n)}$ as general as
possible. In massless QCD two kinds of IR singularities exist: {\it (i)} soft singularities
which arise when a gluon is emitted with vanishing momentum; {\it (ii)} collinear
singularities which arise when two interacting partons are emitted collinearly.  In order to
define infrared finite jet cross sections we have to require finite limits whenever momenta
in the final state belong to either {\it (i)} or {\it (ii)}.  Given a set of functions
$S^{(n)}_J(p_1,\cdots,p_n;p_\pa,p_\pb)$ (where $(p_\pa,p_\pb)$ denote the momenta of the
initial state), we have to require that a state $(p_1,\cdots,p_j,\cdots,p_n)$ with a soft
particle $p_j\to0$ be indistinguishable from the $n-1$-particle state $(p_1,\cdots,p_n)$:
\begin{subequations}\label{Srel}
\begin{equation}\label{Ssoft}
 \lim_{p_j\to0} S^{(n)}_J(\cdots,p_j,\cdots;p_\pa,p_\pb) = S^{(n-1)}_J(\cdots;p_\pa,p_\pb)\;
\end{equation}
with $p_j$ being dropped in the RHS. In the same way, the $n$-particle state
$(p_1,\cdots,p_i,p_{i+1},\cdots,p_n)$ with two collinear particles, e.g. $p_i\parallel
p_{i+1}$, cannot be distinguished from the $n-1$-particle final state
$(p_1,\cdots,p_i+p_{i+1},\cdots,p_n)$.  The jet function must then fulfill
\begin{equation}\label{Scoll}
 S^{(n)}_J(\cdots,ap,bp,\cdots;p_\pa,p_\pb) = S^{(n-1)}_J(\cdots,(a+b)p,\cdots;p_\pa,p_\pb)
 \;,\quad(a,b>0)\;.
\end{equation}
When an outgoing particle is collinear to an incoming one, say $\pa$, a similar relation
(which can be inferred from Eq.~(\ref{Scoll}) by invoking crossing symmetry) holds:
\begin{equation}\label{Sfact}
 S^{(n)}_J(\cdots,a p_\pa,\cdots;p_\pa,p_\pb) = S^{(n-1)}_J(\cdots;(1-a)p_\pa,p_\pb)
 \;,\quad (0<a<1)\;.
\end{equation}
\end{subequations}
The last equation expresses the property of factorizability of initial state collinear
singularities. Eqs.~(\ref{Scoll}) and (\ref{Sfact}) should be understood as smooth limits
for momenta approaching the collinear configuration.

In our case real emission involves three partons in the final states, one gluon in addition
to the incoming quarks $\pa$ and $\pb$: $\pa\pb\pg$.  As indicated in Fig.~\ref{f:ab123}, we
label the outgoing partons in our process $\pa\pb\to123$ by $1=$ (quark $\pa$), $2=$ (quark
$\pb$), and $3=$ (gluon $\pg$).  Let us start by listing the possible IR singular
configurations. Only the emission of gluon 3 with vanishing momentum gives rise to soft
singularities. Collinear singularities arise in collinear emissions of partons that couple
directly to each other, i.e., have a common vertex.  This is the case for pairs of gluons,
for quarks and gluons, and for identical incoming and outgoing quarks (or quarks and
antiquarks). Therefore, the list of all possible collinear singular configurations, in an
obvious notation, reads as follows:
\begin{subequations}\label{collconf}
\begin{align}\label{collconf1}
&& &\pa\parallel1\;,&&\pa\parallel3\;,&&1\parallel3\;,&&\\ \label{collconf2}
&& &\pb\parallel2\;,&&\pb\parallel3\;,&&2\parallel3\;.&&
\end{align}
\end{subequations}
It is important to note that, in the kinematic regime we are considering, configurations in
which quark 2 is emitted outside the fragmentation region of quark $\pb$ are strongly
suppressed, i.e., quark 2 never belongs to the jet produced in the forward direction of quark
$\pa$. To see this we note that as long as quark $\pb$ is deflected at small angles by gluon
exchange, the amplitude contains a $1/t$ factor coming from the gluon propagator. Suppose,
on the other hand, quark $\pb$ deflected with an angle large enough to enter the central
region or even the fragmentation region of quark $\pa$ (which includes the jet region). If
the scattering happens by gluon exchange, the propagator of the latter is of order
$1/|u|\simeq 1/s$, providing a suppression factor $\sim t/s$ compared to the small angle
case.  There is also the possibility of quark exchange that involves a $1/t$ propagator, but
in this case the spin of the quark introduces a new suppression factor $\sim\sqrt{-t/s}$
compared to the gluon exchange.  Therefore, we can safely neglect the configurations in
which quark 2 enters the jet, and only particles 1 and 3 play a role in building up the jet.

To become more specific, we change the argument structure of the jet distribution functions
$S^{(n=3)}_J$ and introduce, as independent variables, $\pp_1, \pp_3, p_3^+, p_\pa^+$:
\begin{equation}\label{defS}
 \Sj3\big(\pp_1,\pp_3,\frac{p_3^+}{p_H^+};\frac{p_\pa^+}{p_H^+}\big) 
 \equiv \Sj3(\kk',\kk-\kk',xz;x)
 \dug S^{(3)}_J(p_1,p_2,p_3;p_\pa,p_\pb)\;. 
\end{equation}
The soft IR constraint~(\ref{Srel}a) of $S_J^{(3)}$ applies only to soft gluon emission and,
since quark 2 does not participate in the jet, the collinear conditions (\ref{Srel}b,c)
apply only to the configurations listed in~(\ref{collconf1}).  The corresponding relations
for $\Sj3$ read:
\begin{subequations}\label{relS}
\begin{align}\label{softS}
 &\text{3 soft}: & \Sj3(\pp,\mbf0,0;x) &= \Sj2(\pp;x) &&\\ \label{collS}
 &1\parallel3: & \Sj3(a\pp,b\pp,\xi;x) &= \Sj2((a+b)\pp;x) && \\ \label{factS1}
 &\pa\parallel1: & \Sj3(\mbf0,\pp,\xi;x) &= \Sj2(\pp;\xi) && \\ \label{factS2}
 &\pa\parallel3: & \Sj3(\pp,\mbf0,\xi;x) &= \Sj2(\pp;x-\xi) \;. &&
\end{align}
\end{subequations}
In the following sections these relations will be used when extracting the divergencies of
the real emission. It is crucial that the singular contributions generated by the real
corrections are proportional to the LO cross section: only in this case cancellations with
the virtual singularities can occur, and the factorization of the collinear singularities
into the PDFs can be performed consistently.  Therefore, the reduction $\Sj3\to\Sj2$ in the
IR singular configurations contained in Eqs.~(\ref{relS}) is a necessary prerequisite.

\subsection{Phase space splitting and the master formula\label{s:pss}}

Before embarking in the analysis of the one-loop real corrections, let us divide the phase
space of the outgoing gluon and present the 'master formula' that we are going to make use
of. We have already pointed out that the $\log s$ term arises from the configurations with
gluon 3 emitted in the central region, while the gluon in the fragmentation region of $\pb$
should mainly contribute to the impact factor correction $h_\pb^{(1)}$ whereas the gluon in
the fragmentation region of $\pq$ should provide the jet vertex $V_\pq^{(1)}$ and the PDF
corrections $f_\pq^{(1)}$. It is easy to define a rapidity cut that separates the two
fragmentation regions: we perform a boost into the positive z-direction which takes us into
the partonic center of mass system (PCMF). It shifts rapidities while transverse energies and
azimuthal angles are preserved.  Consequently, a four momentum $p^{\mu}$ is transformed into
$p'{}^{\mu}$ with
\begin{equation}\label{partonicCM}
 p'{}^\mu=(p'{}^+,p'{}^-,\pp') = (\esp{\Delta y}p^+,\esp{-\Delta y}p^-,\pp)\;,
 \quad\Delta y = \frac12\log\frac1x\;,
\end{equation}
and rapidity is shifted according to $y'=y+\Delta y$.  In the PCMF, we define the cut as the
rapidity center $y'\cut=0$ which corresponds to $y\cut=-\half\log\frac1x$. In the Sudakov
parameterization~(\ref{k}), the form of the cut is very simple:
\begin{equation}\label{cut}
 y_3=y'\cut=0 \quad\iff\quad w\cut=z\cut=\frac{E_3}{\sqrt{xs}}\;.
\end{equation}
Correspondingly, our rapidity phase space is divided into the ``upper half region''
(negative rapidity: $y_3'<0$ or $w >\frac{E_3}{\sqrt{xs}}$) which contains the fragmentation
region of quark $\pb$ and half of the central region, and the ``lower half region'' (positive
rapidity: $y_3'>0$ or $z> \frac{E_3}{\sqrt{xs}}$) which contains the other half of the
central region and the fragmentation of parton $\pq$ and contributes to the jet vertex.

Finally, we need the partonic differential cross section for the $\pb\pq\to123$ process.
They have been computed in Refs.~\cite{Ci98,CiCo98}.  In the high energy regime, where we
neglect terms suppressed by powers of $s$, the form of the partonic differential cross
section turns out to be quite simple when restricted to one of the two halves of the phase
space, $y_3'<0$ or $y_3'>0$. For the ``lower half region'' $y_3'>0$ the cross section can be
cast into the general form
\begin{equation}\label{realpart}
 \dif\hat\sigma_{\pb\pa\to{\it fin}} = \hz_\pb(\kk) F_{\it fin}(\kk,\kk',z) \hz_\pa(\kk')
  \;\dk\, \dk'\, \dif z\;,\quad(z>z\cut)\;,
\end{equation}
where the function $F$ depends on the particular final state.  In particular, for
quark-quark scattering, we have
\begin{align}\label{Fqqg}
 F_{\pq\pq\pg}(\kk,\kk',z) &= \frac{\as}{2\pi} \frac{\PP_{\pg\pq}(z,\e)}{\pi_\e}
  \frac{1}{\qq^2(\qq-z\kk)^2}
  \left[ C_F \, z^2 \kk'\,^2+C_A\, (1-z)\, \qq\cdot (\qq-z\kk)\right]\;,\\ \label{Pgq}
 &\PP_{\pg\pq}(z,\e) = \frac{1+(1-z)^2+\e z^2}{z}\;,\quad
 \pi_\e = \pi^{1+\e}\Gamma(1-\e)\mu^{2\e}\;,
\end{align}
where $\qq=\kk-\kk'$ is the gluon transverse momentum and $\PP_{\pg\pq}(z,\e)$ is --- apart
from a missing $C_F$ factor --- the real part of the $\pq\to\pg$ splitting function in
$4+2\e$ dimensions.  In the ``upper half region'' $y_3'<0$, the same relation
(\ref{realpart}) holds with the replacements
\begin{equation}\label{replacement}
 \kk\to-\kk'\;,\quad \kk'\to-\kk\;,\quad \qq\to\qq\;,\quad z\to w\;,
\end{equation}
except for the impact factor which retain their form.  These 'master formulae' (\ref{Fqqg})
- (\ref{replacement}) will be used in the following in order to find the real corrections to
the quark-initiated jet vertex.

\subsection{Real corrections to the upper quark impact factor\label{s:ifc}}

We begin by computing the contribution to the jet cross section given by the ``upper half
region'' $y_3'<0$ which gives rise to the NLO impact factor of the upper quark. The starting
formula is derived from Eq.~(\ref{pSf}), using Eq.~(\ref{defS}) for the jet distribution
(with $z=E_3^2/wxs$ to good accuracy) and Eqs.~(\ref{realpart}) and (\ref{Fqqg}) (with the
replacements~(\ref{replacement})) for the partonic cross section:
\begin{align}\label{dJneg}
 \dJ{^{(y_3'<0)}} &= \frac{\as}{2\pi}\int\dk\,\dk'\;\hz_\pb(\kk)\hz_\pq(\kk')
 \int_{w\cut}^1\dif w\;\frac{\PP_{\pg\pq}(w,\e)}{\pi_\e}\frac{1}{\qq^2(\qq+w\kk')^2}
 \times \\ \nonumber
 &\quad\left[C_F w^2\kk^2+C_A (1-w) \qq\cdot (\qq+w\kk')\right]
 \int_0^1\dif x\; \Sj3\big(\kk',\qq,\frac{E_3^2}{wxs}x;x\big) \fz_\pq(x)\;.
\end{align}
\begin{figure}[b]
\centering
\includegraphics[height=30mm]{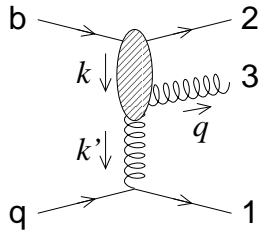}
\caption{\label{f:negative}Structure of the Feynman diagrams contributing to the
  $\pq\pb\to\pq\pb\pg$ process in NLL approximation when the outgoing gluon is emitted with
  negative rapidity $y_3'<0$. The blob represents all possible $\pb\pg^*\to\pb\pg$ QCD
  sub-diagrams.}
\end{figure}
Since $y_3<y_3'<0$, the gluon is emitted very far from the jet region $y\simeq y_J\pm R$,
for which $y_J\sim\log\sqrt s \gg R \sim 1$ (cf.~Eq.~(\ref{logrowth})). Therefore, only
quark 2 can enter the jet, so that in this half of the phase space
\begin{equation}\label{S3neg}
 \Sj3\big(\kk',\qq,\frac{E_3^2}{wxs}x;x\big) = \Sj3\big(\kk',\qq,\frac{E_3^2}{ws};x\big) = \Sj2(\kk';x)
 \;,\quad(w>w\cut)\;.
\end{equation}
It is now possible to factor out from the $w$ and $\kk$ integrals the $\pq$ impact factor
and the jet distribution, which, according to Eq.~(\ref{dV0}), reproduce the LO jet vertex.
We obtain:
\begin{align}\label{UVf}
 \dJ{^{(y_3'<0)}} &= \as\int\dif x \int \dk'\;U(\kk',x)
 V_\pq^{(0)}(\kk',x)\fz_\pq(x), \\ \label{defU}
 U(\kk',x) &\dug \frac{\N C_F}{2\pi}\int\frac{\dk}{\pi_\e\kk^2}\int_{w\cut}^1\frac{\dif w}{w}\;
 \left[1+(1-w)^2+\e w^2\right]
 \frac{C_F w^2\kk^2+C_A (1-w) \qq\cdot (\qq+w\kk')}{\qq^2(\qq+w\kk')^2}\;.
\end{align} 
The computation of the $U$ integral can be done following the calculation of
Ref.~\cite{Ci98}. We repeat here the main steps.  Let us consider separately the two terms
involving different colour constants.

The integrand of the $C_F$ term of $U$ is regular --- actually vanishes --- for $w\to0$ at
fixed $\kk$, so that the lower bound $w\cut$ can be set equal to zero, introducing a
negligible error of order $E^2_3/s$. Changing the transverse integration variable
$\kk\to\qq=\kk -\kk'$ yields
\begin{equation}\label{defUcf}
 U_{C_F} \simeq \frac{\N C_F^2}{2\pi} \int_0^1\dif w\;w\left[1+(1-w)^2+\e w^2\right]
 \int\frac{\dif\qq}{\pi_\e}\;\frac1{\qq^2(\qq+w\kk')^2}\;.
\end{equation}
Until now, we have not mentioned any upper limit of the transverse momentum integrations. We
know that the transverse momentum squared of the outgoing particles are kinematically
limited to some value of the order of $s$.  For our purposes, however, these upper limits
are not important, because the transverse momentum integrals converge in the ultraviolet
region $\kk\to\infty$, and extending the upper limit of integration to infinity causes
negligible errors of order $1/s$:
\begin{equation}\label{UVerror}
 \int_{|\qq|\sim\sqrt s}^\infty\;\frac{\dif\qq}{\qq^4} \propto\int_s^\infty
\frac{\dif t}{t^2} = \frac1s\;.
\end{equation}
The transverse integral in Eq.~(\ref{defUcf}) can then be easily performed,
\begin{equation}\label{tiUcf}
 \int\frac{\dif\qq}{\pi_\e}\;\frac1{\qq^2(\qq+w\kk')^2} = \frac{w^{2\e-2}}{\kk'{}^2}
 \left(\frac{\kk'{}^2}{\mu^2}\right)^\e \frac{\Gamma^2(\e)}{\Gamma(2\e)} =
 \frac{w^{2\e-2}}{\kk'{}^2}\left(\frac{\kk'{}^2}{\mu^2}\right)^\e
 2\left[\frac1\e-\frac{\pi^2}6\e+\ord(\e^2)\right]\;.
\end{equation}
The $\e$ pole reflects the two collinear singularities at $\qq=0$ and $\qq+w\kk'=0$ which
correspond to $\pb\parallel3$ and $2\parallel3$, respectively. At this point also the
$w$-integration can be easily performed. Because of the pre-factor $w^{2\e-2}$ in
Eq.~(\ref{tiUcf}), the $w$-integration turns out to be
divergent%
\footnote{The reader could object that the $w=0$ singularity is an artifact of having set
  $w\cut=0$ which corresponds to having neglected an infinite contribution. However this
  procedure is consistent in that the logic of this calculation is first to perform the Regge
  limit~(\ref{HElimit}) at $\e\neq0$ and only as a last step, after cancellation of the
  divergencies, to perform the physical $\e\to0$ limit.} 
at $w=0$. This is just the soft singularity expected for $p_3\to0$.  The final result for
$U_{C_F}$ is
\begin{equation}\label{Ucf}
 U_{C_F} = \frac{C_F}{\pi}\hz_\pb(\kk')\left(\frac{\kk'{}^2}{\mu^2}\right)^\e
 \left[\frac1{\e^2}-\frac3{2\e}+4-\frac{\pi^2}6\right]
\end{equation}
up to terms $\ord(\e)$.  For the impatient reader we note that the whole $U_{C_F}$ cancels
out when added to the virtual corrections of Sec.~\ref{s:virt}.

The $C_A$ term of $U$ requires special care, because it contributes to the high energy
leading $\log s$ piece of the cross section.  In Sec.~\ref{s:1la} (Eq.~(\ref{central})) we
have presented the leading partonic differential cross section, which is nothing but the
partonic cross section~(\ref{realpart}) with Eq.~(\ref{Fqqg}) evaluated in the $z\to0$ limit
(the central region).  The same consideration holds also here after the replacements
(\ref{replacement}).  In particular we identify the leading term of $U_{C_A}$ with
\begin{align}\label{defUcaL}
 U_{C_A}^{\rm LL} &= \frac{\N C_A C_F}{\pi}\int\frac{\dk}{\pi_\e\kk^2}\;\frac1{(\kk-\kk')^2}
 \int_{w\cut}^1\dif w\;\frac{\vartheta(w,\kk,\kk')}{w} \\ \nonumber
 &= 2\pi\int\dk\;\hz_\pb(\kk)K^{(0,{\rm real})}(\kk,\kk')
 \int_{w\cut}^1\dif w\;\frac{\vartheta(w,\kk,\kk')}{w}\;,
\end{align}
where $K^{(0,{\rm real})}$ is defined in Eq.~(\ref{dK0real}).  The function $\vartheta$ in
the above equation signals that the functional form of the integrand, accurate in the
central region, has to break down somewhere in the fragmentation region of quark $\pb$.
According to the analysis of Ref.~\cite{Ci88}, the emission probability in the splitting
$\pq\to\pq'\pg$ is dynamically suppressed when the emission angle of the gluon $\pg$ is
smaller than that of the quark $\pq'$. In the present case, by comparing the ratios of
transverse to longitudinal components of particles 2 and 3, one expects the active phase
space to be
\begin{equation}\label{angordneg}
 \frac{E_3}{w} > \frac {E_2}{1-w} \quad\iff\quad w < \frac{E_3}{E_2+E_3}\;.
\end{equation}
This has led the author of Ref.~\cite{Ci98} to propose a leading term of the form of
Eq.~(\ref{defUcaL}) with $\vartheta(w,\kk,\kk') \dug \Theta(E_3 - w E_2)$, which matches
Eq.~(\ref{angordneg}) in the low-$w$ region. With this choice,
\begin{align}\label{UcaLciafa}
 U_{C_A}^{\rm LL} &= \int\dk\;\hz_\pb(\kk)K^{(0,{\rm real})}(\kk,\kk')
 \log\frac{\sqrt{xs}}{r(\kk,\kk')}\\ \nonumber
 r(\kk,\kk') &= \max(E_2,E_3)\;.
\end{align}
The remaining part of $U_{C_A}$, i.e., $U_{C_A}-U_{C_A}^{\rm LL}$, is constant in $s$, and
hence it is a NLL contributions; it is given by an integral that, at fixed transverse momenta,
is now finite for $w\cut\to0$. The result is
\begin{equation}\label{UcaNLciafa}
 U_{C_A}^{\rm NLL} = \frac{C_A}{\pi}\hz_\pb(\kk')\left(\frac{\kk'{}^2}{\mu^2}\right)^\e
 \left[-\frac3{4\e}-\frac{\pi^2}3-\frac14\right]\;.
\end{equation}
The $\e$-pole stems from the transverse $\kk$-integration in the neighbourhood of the
singularity at $\kk=0$.

The complete contribution of the ``upper half region'' to the jet cross section can be
conveniently presented if combined with the virtual correction contribution of
Eq.~(\ref{dJvirt}) coming from the $\pb$ impact factor correction $\widetilde\Pi_\pb$:
\begin{subequations}\label{fullnegative}
\begin{align}\nonumber
 &\dJ{^{(y_3'<0)}}+\left.\dJ{^{({\rm virt})}}\right|_{\widetilde\Pi_\pb} =
 \\ \label{LLnegative}
 &\quad=\as\int\dif x\int\dk\,\dk'\;\hz_\pb(\kk) K^{(0,{\rm real})}(\kk,\kk')
 \log\frac{\sqrt{xs}}{\max(E_2,E_3)}V^{(0)}_\pq(\kk',x) \fz_\pq(x) + \\ \label{NLLnegative}
 &\qquad\;\,\as\int\dif x\int\dk\;h^{(1)}_\pb(\kk) V^{(0)}_\pq(\kk,x) \fz_\pq(x)\;,
\end{align}
\end{subequations}
In addition to a LL part, this formula reproduces the first constant term of
Eq.~(\ref{dJ1loop}) (of course, only the term $\pa=\pq$), namely the full one-loop impact
factor correction of the upper quark $\pb$:
\begin{equation}\label{defh1}
 h^{(1)}_\pb(\kk) = \frac{C_A}{\pi}
 \left[\left(-\frac34+\frac{\e}4\right)\frac1\e+\frac{67}{36}-\frac{\pi^2}{12}
 -\frac{5\Nf}{18C_A}\right]\left(\frac{\kk^2}{\mu^2}\right)^\e - b_0\log\frac{\kk^2}{\mu^2}\;.
\end{equation}

The quark impact factor has been calculated in also in Ref.~\cite{FFKP00}, but there a
different definition has been used. In order to explain the relation between the two
approaches, some general remarks on the definition of impact factors and energy scales might
be in place.  It is known that processes involving coloured incoming particles are affected
by collinear singularities that lead to divergent cross sections. These singularities depend
on the type of the incoming particles, and it is therefore natural to associate them with
the process dependent impact factors. In our paper, we follow Ref.~\cite{Ci98,CiCo98}, and
we require {\it the partonic impact factors to include singularities of collinear origin
  only}. Such a prescription may sound somewhat academic, because partonic impact factors
have no phenomenological application and interpretation. However, in view of the jet vertex
which in the present paper represents our main goal, this requirement is very natural.

Namely, at the lower end of the diagram, where the coupling of the reggeized gluon to the
incoming parton is described by the jet vertex (more precisely, by the convolution
$V_\pa\otimes f_\pa$ which we may call ``jet impact factor'') it is mandatory that its
singularity structure matches the one required by collinear factorization. Only in this case
$f_\pa$ can be identified as the usual parton density with one-loop
corrections~(\ref{deff1}), and only with this prescription the remaining jet vertex $V_\pa$
is finite. As we shall see in the next section, the jet distribution function helps to
disentangle this structure from the leading log part in a very natural way, but the basic
requirement is the matching of the singularity structure with the collinear singularities.

In our approach we therefore insist on the collinear properties of any impact factor. It is
apparent from Eq.~(\ref{defh1}) that $h^{(1)}_\pb$ has a simple pole which is the one
expected from the $\pb\parallel2$ collinear divergence --- more precisely, the residue of
the pole reproduces the non singular part of the $\gamma_{\pg\pq}$ anomalous dimension.
This is a consequence of our choice for the LL subtraction needed to define the impact
factor: the angular ordering prescription takes proper account of the whole neighbourhood of
the $\pb\parallel2$ collinear region, and it avoids spurious singularities which potentially
could be present due to the $(\kk-\kk')^2$ denominator in Eq.~(\ref{defUcaL}).
Nevertheless, even with this requirement on the singularity structure there is still some
freedom in the definition of the impact factor --- essentially a factorization-scheme
arbitrariness --- which corresponds to changes in its finite (in $\e$) part. It is possible
to show that the ``good'' collinear properties of the impact factors are preserved, provided
the angular ordering prescription in the leading term is fulfilled in the soft (small
$w,E_3$) region, while the details of the subtraction at finite $w$ or $E_3$ affect only the
finite part.

Returning to the evaluation of leading term, one might think of two different strategies.
First, we could consider the whole $U_{C_A}^{\rm LL}$ term as part of the kernel: this
amounts to perform the $w$-integral in Eq.~(\ref{defUcaL}) and to obtain the LL kernel,
multiplied by a logarithm of the energy. This corresponds to the first line of
Eq.~(\ref{UcaLciafa}), where the energy scale $r$ (the denominator in the argument of the
log), being a function of $\kk$ and $\kk'$, is determined by the particular LL subtraction,
i.e., by $\vartheta$. If we adopt, for instance, the angular ordering prescription in the
whole range of $w$, i.e., $\vartheta(w,\kk,\kk')\dug\Theta(E_3-w (E_2+E_3))$, the energy
scale is $r=E_2+E_3$.  In this case the new impact factor differs from Eq.~(\ref{defh1})
only by a finite piece.

Alternatively , we might chose a particular energy scale, say $\bar{r}$, and then decompose
$U_{C_A}^{\rm LL}$ into a leading term containing the $\log(\sqrt{xs}/\bar{r})$, plus a
next-to-leading term containing the $\log(r/\bar{r})$. The latter term can be expressed, at
relative order $\ord(\as)$, as a multiplicative operator factor:
\begin{align}\label{hHG}
 &(\hz+\as h^{(1)})(1+\as H_L^{(\bar{r})})\left(1+\as K^{(0)}\log\frac{\sqrt{xs}}{\bar{r}}\right)\\
 \label{defHL}
 &H_L^{(\bar{r})}(\kk,\kk') \dug K^{(0)}(\kk,\kk') \log\frac{\bar{r}(\kk,\kk')}{r(\kk,\kk')}\;.
\end{align}
As long as we have chosen the energy scale $\bar{r}$ in such a way that in the limit
$\kk\ll\kk'$ ($\kk-\kk'\ll\kk$) it reduces to $\bar{r}\to|\kk'|\simeq|\kk-\kk'|=E_3$
($\bar{r}\to|\kk|=E_2$), the factor $(1+\as H_L)$ can be safely embodied in the impact
factor term without spoiling its collinear properties, i.e., it changes $h^{(1)}$ only by a
finite (in $\e$) amount.

On the other side, for Regge-motivated scales of the energy like the one proposed in
Ref.~\cite{FaLi98}, the inclusion of the $H_L$ term in the impact factors%
\footnote{In Ref.~\cite{FFKP00}, one can find the expression for the impact factor at the
  Regge-scale $\bar{r}=|\kk'|=E_2$ and the equivalence with our expression $(\hz+\as
  h^{(1)})(1+\as H_L^{(\bar{r})})$ is proven.}
leads to a different infrared behaviour, giving rise even to double poles. In this case, in
our opinion, it will be useful to separate the factor $(1+\as H_L)$ from the impact factor
and eventually to include it into the definition of the NLO BFKL kernel.

\subsection{Real corrections to the jet vertex\label{s:pgr}}

In this section we consider the real corrections in the ``lower half region'' $y_3'>0$,
i.e., the corrections to the jet vertex.  The starting formula is derived from
Eq.~(\ref{pSf}), using Eq.~(\ref{defS}) for the jet distribution and Eqs.~(\ref{realpart})
and~(\ref{Fqqg}) for the partonic cross section:
\begin{align}\label{dJpos}
 \dJ{^{(y_3'>0)}} &= \frac{\as}{2\pi}\N C_F\int\dk\,\dk'\;\hz_\pb(\kk)
 \int_{z\cut}^1\dif z\;\frac{\PP_{\pg\pq}(z,\e)}{\pi_\e}
 \frac{1}{\kk'{}^2\qq^2(\qq-z\kk)^2} \times \\ \nonumber
 &\quad\left[C_F z^2\kk'{}^2+C_A (1-z) \qq\cdot (\qq-z\kk)\right]
 \int_0^1\dif x\; \Sj3\big(\kk',\qq,xz;x\big) \fz_\pq(x)\;,
\end{align}
where we have used Eq.~(\ref{dh0}) in order to write explicitly the kinematic dependence of
the $\pq$ impact factor.

\begin{figure}[hb!]
\centering
\includegraphics[height=30mm]{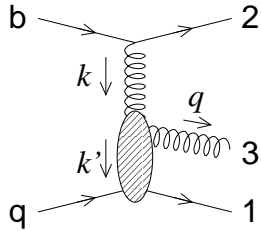}
\caption{\label{f:positive}Structure of the Feynman diagrams contributing to the
  $\pq\pb\to\pq\pb\pg$ process in NLL approximation when the outgoing gluon is emitted with
  positive rapidity $y_3'>0$. The blob represents all possible $\pq\pg^*\to\pq\pg$ QCD
  sub-diagrams.}
\end{figure}
In this phase space region, the quark-quark $\{\pb,2\}$ subsystem is kinematically well
separated from the quark-quark-gluon $\{\pq,1,3\}$ system, and they are connected only via
the exchange of a gluon of momentum $k$ (Fig.~\ref{f:positive}). Therefore, at fixed $k$,
the dynamics of the two subsystems are independent of each other.  Actually, because of the
$\kk$-factorization~\cite{CaCiHa90}, only the transverse momentum $\kk$ and the longitudinal
component $w$ need to be fixed in order to separate the two systems, because the $\{\pb,2\}$
system does not contain final state emissions at large sub-energies, and its dependence on
the longitudinal component $\bar w$ is very weak and can be neglected.  Moreover, the value
of $w$ is constrained by the mass-shell condition of quark 2, so that only $\kk$ is the
relevant variable between the two subsystems. We therefore expect a ``pure''
$\kk$-factorization, where the $\{\pb,2\}$ coupling with the exchanged gluon is described
simply by the $\hz_\pb(\kk)$ impact factor.  In order to find the expression for the
$\{\pq,1,3\}$ system and its coupling to the gluon (and to the hadron $H$), we fix the
transverse momentum $\kk$, remove the $\pb$ impact factor and study the remaining part of
the jet cross section Eq.~(\ref{dJpos}).

First of all, it is important to understand the structure of the singularities in
Eq.~(\ref{dJpos}). The integrand contains three singular points in the $\kk'$-integration,
namely the zeroes of the denominator $\kk'=0$, $\qq=0$, and $\qq-z\kk=0$. These points
correspond to the collinear configurations $\pq\parallel1$, $\pq\parallel3$ and
$1\parallel3$, respectively.  Moreover, there is a potential soft singularity hidden in the $1/z$
pole of the ``splitting function'' $\PP_{\pg\pq}$. The numerators in the square brackets of
Eq.~(\ref{dJpos}) soften some of those singularities, but this happens differently for the
$C_F$ and the $C_A$ parts. Therefore, we consider these two terms separately.

\subsubsection{$C_F$ term}

The $C_F$ term, owing to the $\kk'{}^2$ factor in the numerator, has no $\pq\parallel1$
collinear singularity. Due to the factor $z^2$ in the numerator, the $z$-integrand is no
longer singular at $z=0$ and, along the same line of arguments as given in Sec.~\ref{s:ifc},
we can shift the value of $z\cut\to0$.  Dropping the ($y_3'>0$) label, the $C_F$ part of
Eq.~(\ref{dJpos}) is given by
\begin{align}\label{CF1}
 \dJ{_{C_F}} &= \as\frac{\N C_F^2}{2\pi}\int\dk\;\hz_\pb(\kk) \int_0^1\dif x\;\fz_\pq(x)
 \times \\ \nonumber
&\quad\int_0^1\dif z\;z^2\PP_{\pg\pq}(z,\e)
 \int\frac{\dk'}{\pi_\e}\frac1{\qq^2(\qq-z\kk)^2}\Sj{3}(\kk',\qq,xz;x)\;.
\end{align}
It is convenient to rescale the gluon transverse momentum by setting $\qq\ugd z\lb$, and to
use $\lb$ as integration variable by substituting $\kk'=\kk-z\lb$, so that
$\qq-z\kk=z(\lb-\kk)$. Next we perform a simple fraction decomposition in order to separate
the initial ($i$) state ($\pq\parallel3\iff\lb=0$) and final ($f$) state
($1\parallel3\iff\lb-\kk=0$) collinear singularities:
\begin{equation}\label{fract}
 \frac1{\lb^2(\lb-\kk)^2} = \frac1{\lb^2+(\lb-\kk)^2}\left[\frac1{\lb^2}+\frac1{(\lb-\kk)^2}
 \right]\;.
\end{equation}

Beginning with the {\bf final state} ($f$) collinear singularity, in terms of the new
variables the $C_F$ contribution to the jet cross section can be rewritten in the form
\begin{equation} \label{dJCFf}
 \dJ{^{f}_{C_F}} \dug \as\int_0^1\frac{\dif z}{z^{1-2\e}}
 \int\frac{\dif\lb}{\pi_\e(\lb-\kk)^2} I(z,\lb) =
 \dJ{^{f,{\rm soft}}_{C_F}}+\dJ{^{f,{\rm coll}}_{C_F}}+\dJ{^{f,{\rm finite}}_{C_F}}\;,
\end{equation}
which is particularly suitable for the analytic extraction of the divergencies: the RHS
contains three pieces, {\it 1)} the soft divergence, {\it 2)} the collinear divergence, and
{\it 3)} a finite part.  The explicit expression of the integrand $I$ in Eq.~(\ref{dJCFf})
is
\begin{equation}\label{idJCFf}
 I(z,\lb) \dug \frac{C_F}{2\pi} z\PP_{\pg\pq}(z,\e)\int\dk\;\hz_\pb(\kk)
 \frac{\N C_F}{\lb^2+(\lb-\kk)^2}
 \int_0^1 \dif x \, \Sj{3}(\kk-z\lb,z\lb,xz;x) \fz_\pq(x)\;.
\end{equation}

The soft term in Eq.~(\ref{dJCFf}) is defined by evaluating the integrand in the soft limit
$z\to0$. In this limit, the jet distribution can be simplified by means of
Eq.~(\ref{softS}), and leads to the constraint $\kk^2=E_J^2$. One obtains
\begin{align} \label{CFfsoft}
\dJ{^{f,{\rm soft}}_{C_F}} &\dug \as\int_0^1\frac{\dif z}{z^{1-2\e}}
 \int\frac{\dif\lb}{\pi_\e(\lb-\kk)^2} I(0,\lb) \\ \nonumber
&= \as\frac{C_F}{\pi}\int\dk\;\hz_\pb(\kk) \int_0^1 \frac{\dif z}{z^{1-2\e}}
\int \frac{\dif \lb}{\pi_\e(\lb-\kk)^2}\frac{\N C_F}{\lb^2+(\lb-\kk)^2}
\int_0^1 \dif x \, \Sj{2}(\kk; x) \fz_\pq(x) \\ \nonumber
&= \as\frac{C_F}{\pi} \left[\frac{1}{2\e^2} -\frac{\pi^2}{12} +\ord(\e)  \right]
 \left(\frac{E_J^2}{\mu^2} \right)^\e
 \int\dk\int_0^1\dif x\;\hz_\pb(\kk) V^{(0)}_\pq(\kk,x)\fz_\pq(x)\;,
\end{align}
where, in the final result, we have collected some factors in such a way that they reproduce
the LO jet vertex~(\ref{dV0}). It can easily be seen that we have recovered the LO structure
of the factorization formula. The additional divergent factor exhibits single as well as
double poles, because our definition of the soft part includes also the region where
collinear and soft singularities merge.

The pure collinear singularity can be isolated by evaluating the integrand~(\ref{idJCFf}) in
the collinear limit $\lb=\kk$, after having subtracted the soft term $(\lb=\kk,z=0)$.  The
resulting expression is clearly regular in the soft limit ($z \to 0$) and therefore contains
a simple collinear pole.  An UV cutoff $\Lambda$ is introduced since the residue at the
collinear limit is no more integrable in the UV region.  Thanks to Eq.~(\ref{collS}) the jet
distribution simplifies to $\Sj2$:
\begin{align} \label{CFfcoll}
 \dJ{^{f,{\rm coll}}_{C_F}} &\dug \as\int_0^1 \frac{\dif z}{z^{1-2\e}}
 \int \frac{\dif \lb}{\pi_\e(\lb-\kk)^2} \left[ I(z,\kk)-I(0,\kk) \right]
 \Theta(\Lambda^2-(\lb-\kk)^2) \\ \nonumber
&= \as\frac{C_F}{2\pi}\int\dk\;\hz_\pb(\kk) \frac{\N C_F}{\kk^2}
 \int_0^1 \frac{\dif z}{z^{1-2\e}}\left[ z \PP_{\pg\pq}(z,\e) -2 \right] \times \\ \nonumber
 &\quad \int\frac{\dif \lb}{\pi_\e(\lb-\kk)^2}\Theta(\Lambda^2-(\lb-\kk)^2)
 \int_0^1 \dif x \, \Sj{2}(\kk; x) \fz_\pq(x) \\ \nonumber
&= \as\frac{C_F}{\pi}\left[-\frac{3}{4\e}\left(\frac{\Lambda^2}{\mu^2}\right)^\e+2+\ord(\e)\right]
 \int_0^1\dif x\int \dk\;\hz_\pb(\kk) V^{(0)}_\pq(\kk,x)\fz_\pq(x) \;.
\end{align}

The remaining part is regular in the $\e\to0$ limit and defines the finite term:
\begin{equation}\label{CFffinite}
 \dJ{^{f,{\rm finite}}_{C_F}} \dug \as\int_0^1 \frac{\dif z}{z}
 \int\frac{\dif \lb}{\pi(\lb-\kk)^2}\big[ I(z,\lb)-I(0,\lb)
 -\big( I(z,\kk)-I(0,\kk)\big)\Theta(\Lambda^2-(\lb-\kk)^2) \big] 
\end{equation}
This result (with explicit expressions for $I(z,\kk)$ etc.) will be used in our final
formula.

Next we consider the term ($i$) with the {\bf initial state} collinear singularity. We can
write, in the same way as before,
\begin{equation} \label{dJCFi}
 \dJ{^{i}_{C_F}} \dug
 \as\int_0^1 \frac{\dif z}{z^{1-2\e}} \int \frac{\dif \lb}{\pi_\e \lb^2} I(z,\lb)
 = \dJ{^{i,{\rm soft}}_{C_F}}+\dJ{^{i,{\rm coll}}_{C_F}}+\dJ{^{i,{\rm finite}}_{C_F}} \;,
\end{equation}
where $I$ is given by Eq.~(\ref{idJCFf}).  One sees immediately that the soft contribution
is exactly the same as for the ($f$) term:
\begin{equation} \label{CFisoft}
 \dJ{^{i,{\rm soft}}_{C_F}} = \dJ{^{f,{\rm soft}}_{C_F}}\; .
\end{equation}
As to the collinear piece, we note that in the collinear limit $\lb=0$ the jet distribution
reduces (by applying Eq.~(\ref{factS2})) to $\Sj2$, and one gets%
\footnote{The UV cutoff $\Lambda$ for the initial state collinear singularity is in
  principle independent of the one adopted in the final state collinear term of
  Eq.~(\ref{CFfcoll}). We use the same cutoff for all collinear subtractions in view of its
  identification with the factorization scale $\Lambda=\mu_F$.}
\begin{align}\label{defdJCFicoll}
 \dJ{^{i,{\rm coll}}_{C_F}} &\dug \as\int_0^1 \frac{\dif z}{z^{1-2\e}}
 \int \frac{\dif \lb}{\pi_\e \lb^2} \left[ I(z,\mbf{0})-I(0,\mbf{0}) \right]
 \Theta(\Lambda^2-\lb^2) \\ \nonumber
&= \as\frac{C_F}{2 \pi}\int \dk\;\hz_\pb(\kk)\frac{\N C_F}{\kk^2}
 \int\frac{\dif \lb}{\pi_\e \lb^2} \Theta(\Lambda^2-\lb^2)
 \int_0^1 \dif x \; \fz_\pq(x) \times \\ \nonumber
&\quad\int_0^1\frac{\dif z}{z^{1-2\e}}\left[z\PP_{\pg\pq}(z,\e)\Sj2(\kk;x(1-z))-2\Sj2(\kk;x)\right]\;.
\end{align}
With the change of variable $z \to 1-z$, and performing the $\lb$ integration, we obtain
\begin{align}\label{dJCFic1}
 \dJ{^{i,{\rm coll}}_{C_F}} &= \as\frac{C_F}{2 \pi \e}\left(\frac{\Lambda^2}{\mu^2}\right)^\e
 \int\dk\;\hz_\pb(\kk)\hz_\pq(\kk)\int_0^1\dif x\;\fz_\pq(x)\times \\ \nonumber
&\quad\int_0^1\dif z\;\frac1{\left[(1-z)^{1-2\e}\right]_+}
 \left[1+z^2+\e(1-z)^2 \right]\Sj2(\kk;xz)\;,
\end{align}
where the standard $()_+$ regularization has been used.  To separate singular from finite
pieces one has to perform an $\e$-expansion inside the $z$-integral.  In order to cast the
collinear singularity into the standard form, we introduce the full
4-dimensional Altarelli-Parisi $\pq\to\pq$ splitting function%
\footnote{The appearance of a $\pq\to\pq$ splitting function is simply due to the relation
$P_{\pg\pq}(1-z) = P_{\pq\pq}(z)$ valid for $z<1$.}
\begin{equation}\label{Pqq4}
 P_{\pq\pq}(z) = C_F \left(\frac{1+z^2}{1-z}\right)_+ = 
 C_F \left[\frac{1+z^2}{(1-z)_+} + \frac32\delta(1-z)\right].
\end{equation}
We then can write
\begin{align} \label{CFicoll}
 \dJ{^{i,{\rm coll}}_{C_F}} &=\as\int\dk\;\hz_\pb(\kk)\int_0^1\dif x\;\fz_\pq(x)
 \Biggl\{\frac{C_F}{\pi}\left(-\frac{3}{4\e}\right)\left(\frac{\Lambda^2}{\mu^2}\right)^\e
 V^{(0)}_\pq(\kk,x) + \\ \nonumber
&\qquad\int_0^1\dif z\;V^{(0)}_\pq(\kk,xz)\left[\frac{1}{\e}
 \left(\frac{\Lambda^2}{\mu^2}\right)^\e\frac{P_{\pq\pq}(z)}{2\pi}+\frac{C_F}{\pi}\frac{1-z}2
 +\frac{C_F}{\pi}\left(\frac{\log(1-z)}{1-z} \right)_+(1+z^2)\right]\Biggr\}\;.
\end{align}
In the last expression the first term is again the LO jet cross section, multiplied by a
singular factor; the second term contains the proper quark corrections to the quark
distribution function, while the remaining pieces are finite in the $\e\to0$ limit.

The last contribution in the $C_F$ part is regular in 4 dimensions and defines another
finite term
\begin{equation} \label{CFifinite}
 \dJ{^{i,{\rm finite}}_{C_F}} \dug \as\int_0^1\frac{\dif z}{z}
 \int  \frac{\dif \lb}{\pi \lb^2}\big[ I(z,\lb)-I(0,\lb)-
 \big( I(z,\mbf{0})-I(0,\mbf{0})\big)\Theta(\Lambda^2-\lb^2) \big]
\end{equation}
and will be used in our final result.

\subsubsection{$C_A$ term}

The term proportional to $C_A$ in Eq.~(\ref{dJpos}) reads
\begin{align}\label{CA1}
 \dJ{_{C_A}} &= \as\frac{C_A}{2\pi}\N C_F\int\dif\kk\;\hz_\pb(\kk)\int_0^1\dif x\;\fz_\pq(x)
  \times \\ \nonumber
 &\quad\int_{z\cut}^1\dif z\;(1-z)\frac{\PP_{\pg\pq}(z,\e)}{\pi_\e}
 \int\frac{\dif\kk'}{\kk'{}^2}\;
 \frac{\qq\cdot(\qq-z\kk)}{\qq^2(\qq-z\kk)^2}\Sj3(\kk',\qq,xz;x) \\ \nonumber
 &= \dJ{_{C_A}^{\rm coll}} + \dJ{_{C_A}^{\rm LL}} + \dJ{_{C_A}^{\rm const}}\;.
\end{align}
It shows a $\pq\parallel1$ collinear singularity corresponding to the $\kk'=0$ pole. Because
of the numerator, the $\kk'$-integration is not really singular at $\qq=0$ or at
$\qq-z\kk=0$, except for $z\to0$. In fact, in the high energy limit~(\ref{HElimit}), $\qq$
fixed and $z\ll1$ correspond to gluon 3 being in the central region, where we have
\begin{equation}\label{limcentr}
 \frac{C_A}{2\pi}(1-z)\frac{\PP_{\pg\pq}(z,\e)}{\pi_\e}
 \frac{\qq\cdot(\qq-z\kk)}{\qq^2(\qq-z\kk)^2}\quad\stackrel{z\to0}{\longrightarrow}
\quad \frac{C_A}{\pi}\frac1{\pi_\e\qq^2}\,\frac1z = K^{(0,{\rm real})}(\kk,\kk')\frac1z\;.
\end{equation}
This is exactly the expression entering the differential partonic cross
section~(\ref{central}) in the central region that provides the LL contribution. In other
words: Eq.~(\ref{CA1}) contains the $\pq\parallel1$ collinear singularity in the whole
$z$-range; for finite values of $z$ neither the $\pq\parallel3$ nor the $1\parallel3$
collinear singularities are really present. However, in the (gluon) central region $z\ll1$
their ``collinear denominators'' degenerate, providing the soft singular real part of the
leading kernel $\sim1/\qq^2$.

The jet distribution functions will become essential in disentangling the collinear
singularities, the soft singularities, and the leading $\log s$ pieces. The basic mechanism
can be understood as follows:
\begin{itemize}
\item When the outgoing quark 1 is in the collinear region of the incoming $\pq$, i.e.,
  $y_1\to\infty$, quark 1 cannot enter the jet; only gluon 3 can thus be the jet, $y_3$ is
  fixed and no logarithm of the energy can arise due to the lack of evolution in the gluon
  rapidity. No other singular configuration is found for the quark when $J=\{3\}$.
\item In the composite jet configuration, i.e., $J=\{1,3\}$, the gluon rapidity is bounded
  within a small range of values, and also in this case no $\log s$ can arise. There could
  be a singularity for vanishing gluon 3 momentum: even if the $1\parallel3$ collinear
  singularity is absent, we have seen that, at very low $z$, a soft singular integrand
  arises. However, the divergence is prevented by the jet cone boundary, which causes a
  shrinkage of the domain of integration $\sim z^2$ for $z\to0$ and thus compensates the
  growth of the integrand.
\item The one-quark jet configuration $J=\{1\}$ allows the gluon to span the whole phase
  space, apart, of course, from the jet region itself. The LL term arises from gluon
  configurations in the central region. But also here, like in the negative rapidity region
  discussed in Sec.~\ref{s:ifc}, it is crucial to understand to what extent the differential
  cross section provides a leading contribution. It turns out that the coherence of QCD
  radiation suppresses the emission probability for gluon 3 rapidity $y_3$ being larger than
  the rapidity $y_1$ of the outgoing quark 1, and an angular ordering prescription similar
  to that of Eq.~(\ref{angordneg}) holds. This will provide the final form of the leading
  term, i.e., the appropriate scale of the energy and, as a consequence, a finite and
  definite expression for the one-loop jet vertex correction.
\end{itemize}

As a first step we isolate, in Eq.~(\ref{CA1}), the initial state $\pq\parallel1$ collinear
singular contribution and define, like in the $C_F$-term analysis, the collinear term by
setting $\kk'=0$ (except in the $1/\kk'{}^2$ pole), and by introducing an UV cutoff:
\begin{align}\label{defCAcoll}
 \dJ{_{C_A}^{\rm coll}} &\dug \as\frac{C_A}{2\pi}\N C_F\int\dif\kk\;\hz_\pb(\kk)\int_0^1\dif x
 \;\fz_\pq(x) \\ \nonumber
&\quad \int_{z\cut}^1\dif z\;\PP_{\pg\pq}(z,\e)\int\frac{\dif\kk'}{\pi_\e\kk'{}^2}
 \frac{\Theta(\Lambda^2-\kk'{}^2)}{\kk^2}\Sj3(\mbf0,\kk,xz;x)\;.
\end{align}
In this expression the jet distribution, because of Eq.~(\ref{factS1}), reduces to
$\Sj2(\kk;xz)$. By including the $\N C_A$ constant factors and the $1/\kk^2$ pole, we
reconstruct the {\em gluon-initiated} LO jet vertex (see Eqs.~(\ref{dh0}) and (\ref{dV0})).
Note also that
\begin{equation}\label{Pgq4}
 C_F\PP_{\pg\pq}(z,\e) = C_F\frac{1+(1-z)^2}z+C_F\,\e z = P_{\pg\pq}(z)+C_F\,\e z\;,
\end{equation}
where $P_{\pg\pq}(z)$ is the 4-dimensional $\pq\to\pg$ splitting function. It will be used
to define the quark correction to the {\em gluon} distribution function, and the lower bound
$z\cut$ can be set equal to zero (up to an error of the order $t/s$).  The transverse
integral is easily performed, and we obtain
\begin{equation} \label{CAcoll}
 \dJ{_{C_A}^{\rm coll}} = \as\int_0^1\dif x\int\dk\;\hz_\pb(\kk)\int_0^1\dif z\;
 V^{(0)}_\pg(\kk,xz)\fz_\pq(x) \left[\frac{1}{\e}\left(\frac{\Lambda^2}{\mu^2}\right)^\e 
 \frac{P_{\pg\pq}(z)}{2\pi}+\frac{C_F}{\pi}\frac{z}2\right]+\ord(\e)\;.
\end{equation}

By subtracting the collinear term~(\ref{defCAcoll}) from the total $C_A$ term~(\ref{CA1}),
we obtain an expression, that is finite in the $\e\to0$ limit, and whose value increases
logarithmically with $s$ because of the $\dif z/z$ integration with lower bound $z\cut\sim
s^{-1/2}$. As already mentioned, it is extremely important to understand the extension of
the phase space contributing to the cross section at NLL accuracy, i.e., by including also
the constant terms. That is actually crucial in the small-$z$ region where the integrand
develops a soft singularity which, if not properly absorbed in the real part of the kernel,
could give rise to spurious divergencies spoiling the whole procedure.
  
It is instructive to review how the mechanism of suppression of the differential cross
section sets in when the gluon is emitted at an angle smaller than that of the quark. At
fixed quark momentum $p_1$, i.e., at fixed transverse momentum $\kk'\neq0$ and longitudinal
momentum $p_1^3 \simeq (1-z)x\sqrt{s}/2$ --- one can imagine $J=\{1\}$ fixed by the jet
condition ---, and at fixed gluon transverse energy $E_3=|\qq|$ and longitudinal momentum
$p_3^3 \simeq xz\sqrt{s}/2$, we perform an azimuthal average of the subtracted $C_A$
differential cross section with respect to the angle $\phi_3$ of the gluon. The collinear
subtraction actually does not contribute, because it refers to configuration with $\kk'=0$.
If in this averaging procedure we neglect the variation of $\hz_\pb(\kk)$, it is
sufficient to calculate the azimuthal average of the factor
\begin{equation}\label{average}
 \left\langle\frac{(1-z)\qq\cdot(\qq-z\kk)}{\qq^2(\qq-z\kk)^2}\right\rangle_{\phi_3}
 = \frac1{\qq^2}\Theta(E_3-z(E_1+E_3))\;.
\end{equation}
This relation is exact and clearly shows that, outside the angular ordered region
\begin{equation}\label{angordpos}
 \frac{E_3}{z} > \frac {E_1}{1-z} \quad\iff\quad \theta_3 > \theta_1 
 \quad\iff\quad y_3 < y_1\;,
\end{equation}
there is no contribution to the cross section. In practice, by taking into account the
variation of $\hz_\pb(\kk)$ during the averaging procedure, we have to replace ``no
contribution'' by ``suppression''. Note however that, in the limit $\qq\to0$ (which includes
the soft region), the variation of $\kk'$ goes to zero as well, so that
Eq.~(\ref{angordpos}) is really an accurate statement in the ``dangerous'' part of the phase
space. Moreover, Eq.~(\ref{average}) shows that the $1/\qq^2$ kinematic dependence of the LL
kernel governs the differential cross section up to the very end of the angular boundary.
Therefore, we define the LL contribution in the ``lower half region'' $y_3'>0$ by
\begin{align} \label{CAkern}
 \dJ{_{C_A}^{\rm LL}} &\dug \as\int\dk\;\hz_b(\kk)\int\dk'\;
 \frac{C_A}{\pi}\frac1{\pi_\e \qq^2}\hz_\pq(\kk') \times \\ \nonumber
&\quad\int_{z\cut}^1\frac{\dif z}{z}\;
 \Theta(E_3-z(E_1+E_3))\int_0^1\dif x\;\Sj{2}(\kk', x)\fz_\pq(x) \\ \nonumber
&= \int_0^1\dif x\int\dk\int\dif\kk'\;\hz_\pb(\kk) K^{(0,{\rm real})}(\kk,\kk')
 \log{\frac{\sqrt{xs}}{E_J+E_3}} V^{(0)}_\pq(\kk',x) \fz_\pq(x)\;,
\end{align}
where we have imposed the $J=\{1\}$ jet condition.

The remaining part of the $C_A$ term is finite in 4 dimensions and constant in energy, so
that we can set $\e=0$ and $z\cut=0$ to define the constant part
\begin{equation} \label{CAconst}
 \dJ{_{C_A}^{\rm const}} \dug \left[\dJ{_{C_A}} - \dJ{_{C_A}^{\rm coll}} - \dJ{_{C_A}^{\rm LL}}
 \right]{\!\!}_{\mscr{\begin{array}{r} z\cut=0  \\ \e=0 \end{array}}}\;.
\end{equation}

\section{The NLO jet vertex: sum of real and virtual corrections\label{s:jv}}

Having completed the calculation of both the virtual and real corrections in the whole phase
space, we are going to collect all partial results and to show that the complete one-loop
jet cross section can naturally be fitted to the form of Eq.~(\ref{dJ1loop}).
Table~\ref{t:albero} summarizes the decomposition of the one-loop jet cross section and
gives the references of the various contributions.

In Sec.~\ref{s:virt} we have presented the virtual contributions to the jet cross section
which, after renormalization of the coupling, assume the form of Eq.~(\ref{dJvirt}). We have
already taken into account the contribution coming from the $\widetilde\Pi_\pb$ impact
factor correction in Sec.~\ref{s:ifc} by combining it with the ``upper half region'' real
contribution in Eq.~(\ref{fullnegative}). The remaining virtual terms can be conveniently
rewritten in the form
\begin{subequations}\label{dJvirtpos}
\begin{align}
 &\left.\dJ{^{({\rm virt})}}\right|_{\omega^{(1)}}+\left.\dJ{^{({\rm virt})}}
 \right|_{\widetilde\Pi_\pq}=\nonumber \\ \label{virtLL}
 &=\as\int\dif x\int\dif\kk\,\dif\kk'\;\hz_\pb(\kk) K^{(0,{\rm virt})}(\kk,\kk')
 \log\frac{xs}{s_0(\kk,\kk')} V^{(0)}_\pq(\kk',x)\fz_\pq(x)+ \\ \label{virtQ}
 &\quad\;\as\left[ \left(\frac{E_J^2}{\mu^2}\right)^\e
 \left(-\frac{1}{\e^2}+\frac{3}{2\e}+\frac{\pi^2}{6}-4\right)\frac{C_F}{\pi}
 +\left(\frac{85}{36}+\frac{\pi^2}{4}\right)\frac{C_A}{\pi}-\frac{5}{18}\frac{\Nf}{\pi}
 -b_0\log\frac{E_J^2}{\mu^2}\right]\times \\ \nonumber
 &\qquad\qquad\int\dif x\int\dif\kk\;\hz_\pb(\kk) V^{(0)}_\pq(\kk,x)\fz_\pq(x)\;,
\end{align}
\end{subequations}
where the virtual kernel $K^{(0,{\rm virt})}$ has been defined in Eq.~(\ref{dK0virt}) and
the coefficient $\widetilde\Pi_\pq$ is given in Eq.~(\ref{renqIF}).  The energy scale $s_0$
in Eq.~(\ref{virtLL}) is constrained to satisfy $s_0(\kk,\kk)=\kk^2=E_J^2$, because the delta
function inside the virtual kernel sets $\kk=\kk'$, and the delta function in the LO jet
vertex sets $\kk'{}^2=E_J^2$. The general form of the energy scale for $\kk'\neq\kk$ will be
fixed in a moment by the real LL contribution.
\begin{table}[b]
 \caption{\label{t:albero}Schematics of the decomposition of real and virtual one-loop
   corrections to $\pq\pb$ scattering and references of the corresponding equations.}
 \centering
\begin{tabular}{|cccc|ccc|ccc|ccc|}
\hline
   \multicolumn{3}{|c|}{virtual} & \multicolumn{10}{c|}{real} \\
\hline \slarga{2.7}{8}
   $\omega^{(1)}$ & $\widetilde\Pi_\pq$ & $\widetilde\Pi_\pb$ &
   \multicolumn{1}{|c}{$y_3'<0$} & \multicolumn{9}{|c|}{$y_3'>0$} \\
\cline{5-13}
   \multicolumn{2}{|c}{\slarga{2}{7}} &
   \multicolumn{2}{c|}{$\underbrace{\qquad\quad\qquad}{}$} &
   \multicolumn{6}{|c|}{$C_F$} & \multicolumn{3}{c|}{$C_A$}\\
\cline{5-13}
   \multicolumn{2}{|c}{\slarga{2}{7}} &
   \multicolumn{1}{r}{$\swarrow$} & \multicolumn{1}{l}{$\searrow$} &
   \multicolumn{3}{|c|}{$f$} &\multicolumn{3}{|c|}{$i$} & \multicolumn{3}{|c|}{} \\
\cline{5-10}
  \pic LL & \pic soft,coll & \pic $h^{(1)}_\pb$ & \pic LL & \slarga{2}{7}
   \pic soft & \pic coll & \pic finite & \pic soft & \pic coll & \pic finite & \pic coll & \pic LL & \pic const \\
   (\ref{virtLL}) & (\ref{virtQ}) & (\ref{NLLnegative}) & \multicolumn{1}{c}{(\ref{LLnegative})} &
   (\ref{CFfsoft}) & (\ref{CFfcoll}) & \multicolumn{1}{c}{(\ref{CFffinite})} &
   (\ref{CFisoft}) & (\ref{CFicoll}) & \multicolumn{1}{c}{(\ref{CFifinite})} &
   (\ref{CAcoll}) & (\ref{CAkern}) & (\ref{CAconst}) \\
\hline
\end{tabular}
\end{table}

We first join the LL real contributions~(\ref{LLnegative}) and (\ref{CAkern}): they show the
same structure and differ only in the logarithmic term. The sum of the two logarithms yields
\begin{equation}\label{logtot}
 \log\frac{\sqrt{xs}}{\max(E_2,E_3)}+\log{\frac{\sqrt{xs}}{E_J+E_3}} = 
 \log\frac{xs}{(E_J+E_3)\max(E_2,E_3)}\;.
\end{equation}
The denominator in the argument of the logarithm defines the energy scale $s_0$. However, as
we have already pointed out, there is some freedom in choosing the LL subtraction and,
correspondingly, the denominators in the $\log s$. We can obtain a more symmetric expression
by defining the LL term in the ``upper half region'' as suggested at the end of
Sec.~\ref{s:ifc}: by replacing the scale $\max(E_2,E_3)$ by $E_2+E_3$. This amounts to using
the same prescription for defining the LL contribution in the ``upper half region'' $y_3'<0$
and in the ``lower half region'' $y_3'<0$. With this choice, the full LL contribution to the
jet cross section, including the virtual correction~(\ref{virtLL}), is
\begin{align}\label{fullLL}
 &\dJ{^{\rm LL}} = \int\dif x \int\dif\kk\,\dif\kk'\; \hz_\pb(\kk) K^{(0)}(\kk,\kk')
 \log\frac{xs}{s_0(\kk,\kk')} V^{(0)}_\pq(\kk',x) \fz_\pq(x)
 \\ \label{energyscale}
 &s_0(\kk,\kk') \dug (|\kk'|+|\qq|)(|\kk|+|\qq|) = (E_J+E_3)(E_2+E_3)\;.
\end{align}
It is straightforward to check that
\begin{subequations}\label{s0lim}
\begin{eqnarray}
 s_0(\kk,\kk') &\to& \kk^2 = E_2^2 \;\qquad\text{for}\qquad{\kk^2 \gg \kk'{}^2}\;,\\
 s_0(\kk,\kk') &\to& \kk'{}^2 = E_J^2 \qquad\text{for}\qquad{\kk^2 \ll \kk'{}^2}\;,\\ \label{s0lim3}
 s_0(\kk,\kk) &=& \kk^2 = E_J^2 \;\qquad\text{for}\qquad{\kk = \kk'}\;.
\end{eqnarray}
\end{subequations}
These constraints are consequences of the QCD coherence effects~\cite{Ci88}.  We identify
Eq.~(\ref{fullLL}) with the first term of Eq.~(\ref{dJ1loop}).

Let us stress that the energy scale in Eq.~(\ref{energyscale}) --- and in general all the
scales satisfying Eqs.~(\ref{s0lim}) --- arises naturally when one requires impact factors
and PDFs to have standard collinear properties and the remaining non-leading-log term
(\ref{CAconst}) to be finite in both the physical $\e\to0$ and high-energy $s\to\infty$
limits. Choosing a scale of the energy outside the class defined by Eqs.~(\ref{s0lim}) while
preserving the above properties, requires the introduction of additional NLL operators (see
Eq.~(\ref{defHL})) which has to be added as multiplicative corrections to the Green's
function.  If, for instance, we adopt $s_0=|\kk||\kk'|$, then the Green's function in
Eq.~(\ref{G1loop}) has to be replaced by
\begin{align}\label{HGH}
 G(xs,\kk,\kk')=(\id+\as H_L)\left[\id+\as K^{(0)}\log\frac{xs}{|\kk||\kk'|}\right](\id+\as H_R)\\
  \label{defHLHR} H_L(\kk,\kk')=-K^{(0)}(\kk,\kk')\log\frac{|\kk|+|\qq|}{|\kk|}\;,\quad 
 H_R(\kk,\kk')=H_L(\kk',\kk)\;.
\end{align}

The second term of Eq.~(\ref{dJ1loop}) has already been obtained in Eq.~(\ref{NLLnegative}).
We remark again that, with the choice of the energy scale~(\ref{energyscale}), the one-loop
impact factor correction $h^{(1)}_\pb$ is no longer given by Eq.~(\ref{defh1}) but differs
by a finite part. However, its actual expression is irrelevant for the jet vertex.

We now consider the sum of all $\e$-divergent contributions that have not been included in
the $h^{(1)}_\pb$ impact factor term~(\ref{NLLnegative}). They can be found in
Eqs.~(\ref{virtQ},\ref{CFfsoft},\ref{CFfcoll},\ref{CFisoft},\ref{CFicoll},\ref{CAcoll}) and
add up to
\begin{align}\label{singular}
 \dJ{^{\rm singular}} =\;&
 \frac{\as}{2\pi}\frac1{\e}\left(\frac{\Lambda^2}{\mu^2}\right)^\e
 \int\dif x\int \dif\kk\;\hz_\pb(\kk) V_\pq^{(0)}(\kk,x)\, 
 \big[P_{\pq\pq} \otimes \fz_\pq\big](x) + \\ \nonumber
&\frac{\as}{2\pi}\frac{1}{\e}\left(\frac{\Lambda^2}{\mu^2}\right)^\e
 \int\dif x\int\dif\kk\;\hz_\pb(\kk) V_\pg^{(0)}(\kk,x)\, 
 \left[P_{\pg\pq} \otimes \fz_\pq\right](x)  + \\ \nonumber
&\as\left[\left(\frac{3}{2}\log\frac{E_J^2}{\Lambda^2}-2\right)\frac{C_F}{\pi}+
 \left(\frac{85}{36}+\frac{\pi^2}{4}\right)\frac{C_A}{\pi}-\frac{5}{18}\frac{\Nf}{\pi}
 -b_0\log\frac{E_J^2}{\mu^2}\right]\times\\ \nonumber &\qquad
 \int\dif x\int\dif\kk\;\hz_\pb(\kk) V_\pq^{(0)}(\kk,x) \fz_\pq(x) + \\ \nonumber
&\as\int\dif x\int\dif\kk\;\hz_\pb(\kk) \int\dif z\;V_\pq^{(0)}(\kk,xz)\fz_\pq(x)\times
 \\ \nonumber &\qquad\left\{\frac{C_F}{\pi}\left[\frac{1-z}{2} +
 \left(\frac{\log(1-z)}{1-z}\right)_+ (1+z^2) \right]+\frac{C_A}{\pi}\frac{z}2\right\}\;.
\end{align}
All double poles have cancelled out. Single poles only appear in connection with the
splitting functions: they are shown in the first two lines of Eq.~(\ref{singular}), and they
contribute to the third term of Eq.~(\ref{dJ1loop}) which contains the PDF one-loop
corrections.  Note however that, since the present analysis has been restricted to the case
of incoming quarks, we have obtained only the quark-initiated corrections to the quark and
gluon PDF, i.e., only the term $\pc=\pq$ in the sum of Eq.~(\ref{deff1}). The
gluon-initiated corrections require an incoming gluon out of hadron $H$ and will be
presented in a forthcoming paper~\cite{BaCoVa02}.  It is also clear that the cutoff
$\Lambda$ can be identified with the factorization scale $\mu_F$.

The last two terms of Eq.~(\ref{singular}) are regular at $\e=0$ and can be combined with
the finite parts~(\ref{CFffinite}), (\ref{CFifinite}) and (\ref{CAconst}). The resulting
expression can be cast into the form
\begin{equation}\label{dJV1}
 \dJ{^{\rm finite}} = \as\int\dif x\int\dif\kk\;\hz_\pb(\kk)V_\pq^{(1)}(\kk,x)\fz_\pq(x)\;, 
\end{equation}
which defines the NLO correction to the quark-initiated jet vertex
\begin{align}\label{V1}
V^{(1)}_\pq(\kk,x) \dug\;&
 \left[\left(\frac{3}{2}\log\frac{E_J^2}{\Lambda^2}-2\right)\frac{C_F}{\pi}+
 \left(\frac{85}{36}+\frac{\pi^2}{4}\right)\frac{C_A}{\pi}-\frac{5}{18}\frac{\Nf}{\pi}
 -b_0\log\frac{E_J^2}{\mu^2}\right] V_\pq^{(0)}(\kk,x)+ \nonumber \\
&\int\dif z\; V_\pq^{(0)}(\kk,xz)\left\{\frac{C_F}{\pi}
 \left[\frac{1-z}{2}+\left(\frac{\log(1-z)}{1-z}\right)_+ (1+z^2) \right] 
 +\frac{C_A}{\pi}\frac{z}{2}\right\}  + \nonumber \\
& \frac{C_A}{\pi}\int\frac{\dif\kk'}{\pi}\int\dif z\left[
 \frac{1}{2} P_{\pq\pq}(z) \left( (1-z)\frac{\qq\cdot(\qq-\kk)}{\qq^2 (\qq-\kk)^2}
 \hz_\pq(\kk') \Sj{3}(\kk',\qq,xz;x)+ \right.\right. \nonumber \\
&\quad-\left.\left.\frac{1}{\kk'{}^2}\Theta(\Lambda^2-\kk'{}^2)V_\pq^{(0)}(\kk,xz) \right)
 -\frac{1}{z\qq^2}\Theta(|\qq|-z(|\qq|+|\kk'|)) V_\pq^{(0)}(\kk',x)\right] + \nonumber \\
&\frac{C_F}{2\pi}\int\dif z\;\frac{1}{(1-z)}_+ (1+z^2) \int\frac{\dif\lb}{\pi\lb^2}
 \biggl[\frac{\N C_F}{\lb^2 +(\lb-\kk)^2}\times\nonumber \\
&\quad\left(\Sj{3}(z \kk+(1-z)\lb,(1-z)(\kk-\lb),x(1-z);x) +\right. \nonumber \\
&\left.\qquad\Sj{3}(\kk-(1-z)\lb,(1-z)\lb,x(1-z);x) \right) + \nonumber \\
&\quad-\Theta(\Lambda^2-\lb^2)\Bigl(V_\pq^{(0)}(\kk,xz)+V_\pq^{(0)}(\kk,x)\Bigr)\biggr]\;.
\end{align}
It clearly depends on the jet definition and on three scales: the energy scale $s_0$ (via
the subtraction of the LL term $\propto1/z$), the factorization scale $\Lambda=\mu_F$ and
the renormalization scale $\mu$.

Eqs.~(\ref{dJV1}) and (\ref{V1}) provide the fourth and last term of Eq.~(\ref{dJ1loop}) and
represent the main result of our study.

\section{Conclusive remarks\label{s:con}}

In this paper we have investigated a particular class of jet final states in the high energy
region. Both the theoretical and phenomenological motivation comes from the interest in the
Regge limit of QCD: the jet production processes - forward jets in $ep$ collisions and
Mueller-Navelet jets in hadron-hadron collisions - have been designed to investigate BFKL
dynamics at present and future colliders. Whereas the BFKL Pomeron is known at LO and NLO, a
complete NLO analysis and comparison with data has not been possible, since neither the
photon impact factor nor the jet vertex have been calculated in NLO. On the other hand,
there is no doubt that at moderately large energies, NLO corrections to the asymptotic LO
behaviour are important for a reliable description of the processes under consideration. It
is the purpose of this paper (and a forthcoming one) to calculate the jet vertex in NLO: in
this first part we have studied the quark-initiated jet vertex, whereas the gluonic
counterpart will be presented in a subsequent paper.

In order to extract the NLO jet vertex we have calculated the cross section of the process
$\text{\it quark}\;\pb+\text{\it quark}\;\pa\to\text{\it quark}\;\pb+X+\text{\it jet}\;$ at
order $\alpha_s^3$: apart from the NLO corrections to the impact factor of quark b and the
contribution to gluon production in the central region, this process provides the NLO
corrections to the quark-initiated vertex $\text{\it quark}\;\pa\to\text{\it jet}$. As an important
theoretical result we have verified that the factorization form~(\ref{FF}) holds: this proof
follows from the fact that we have been able to separate, in the sum of virtual and real
corrections, the collinear singularities which go into the renormalization of the parton
density of the incoming quark $\pa$, and the gluon emission in the central region which is
part of the LO calculation.
         
Another theoretical issue of interest is the dependence upon the scales.  In leading order,
results of the BFKL calculations and the jet vertex are insensitive to both the energy scale
$s_0$ and to the renormalization scale $\mu$. It is only at the NLO level that the
dependence on these scales is being determined.
 
The central result of our calculation is the expression~(\ref{V1}) for the NLO jet vertex.
Using the factorization property~(\ref{FF}), we can use our result also for the upper
incoming quark $\pb$, i.e., for the `symmetric' Mueller-Navelet jet production process
$\pq+\pq\to {\it jet}+X+{\it jet}$.  As the final step, we will have to allow for the
production of an arbitrary number of gluons between the jets which is described by the NLO
BFKL Pomeron. This step will be presented in the companion paper containing the NLO
gluon-initiated jet vertex.

With our expression for the jet vertex we have provided a finite integral that can be
computed numerically, e.g. via Monte-Carlo integration. Such a numerical study requires to
specify the jet algorithm described by the function $\Sj3$: so far we been rather general,
but clearly the numerical results will depend on the choice of the jet algorithm.

\section*{Acknowledgments}

We wish to thank John Collins for helpful discussions during the first stages of this work.

\end{document}